\documentclass[aip,jcp,showpacs,preprint,amsmath,amssymb,floatfix]{revtex4-1}
\usepackage{graphicx}
\usepackage{dcolumn}
\usepackage{bm}
\draft 
\begin{document}
\title{Spectroscopy of the Double Minimum $3\,^3 \Pi_{\Omega}$ Electronic State of $^{39}$K$^{85}$Rb} 
\author{Jayita Banerjee}
\email{jbanerjee@phys.uconn.edu}
\affiliation{Department of Physics, University of Connecticut, Storrs, Connecticut 06269-3046, USA}

\author{David Rahmlow}
\altaffiliation[ Presently at ]{Wyatt Technology Corporation, Santa Barbara, CA 93117-3253}
\affiliation{Department of Physics, University of Connecticut, Storrs, Connecticut 06269-3046, USA}

\author{Ryan Carollo}
\affiliation{Department of Physics, University of Connecticut, Storrs, Connecticut 06269-3046, USA}

\author{Michael Bellos}
\affiliation{Department of Physics, University of Connecticut, Storrs, Connecticut 06269-3046, USA}

\author{Edward E. Eyler}
\affiliation{Department of Physics, University of Connecticut, Storrs, Connecticut 06269-3046, USA}

\author{Phillip L. Gould}
\affiliation{Department of Physics, University of Connecticut, Storrs, Connecticut 06269-3046, USA}

\author{William C. Stwalley}
\affiliation{Department of Physics, University of Connecticut, Storrs, Connecticut 06269-3046, USA}
\date{\today}
\begin{abstract}
We report the observation and analysis of the $3\,^3\Pi_{\Omega}$ double-minimum electronic excited state of $^{39}$K$^{85}$Rb. The spin-orbit components ($0^{+}, 0^{-}, 1$ and 2) of this state are investigated based on potentials developed from the available \emph{ab initio} potential curves. We have assigned the vibrational levels $v'=2-11$ of the $3\,^3\Pi_{1,2}$ potentials and $v'=2-12$ of the $3\,^3\Pi_{0^{+/-}}$ potential. We compare our experimental observations of the $3\,^3\Pi_{\Omega}$ state with predictions based on theoretical potentials. The observations are based on resonance enhanced multiphoton ionization (REMPI) of ultracold KRb in vibrational levels $v''=14-25$ of the $a\,^3\Sigma^+$ state. These \emph{a}-state ultracold molecules are formed by photoassociation of ultracold $^{39}$K and $^{85}$Rb atoms to the 5($1$) state of KRb followed by spontaneous emission to the \emph{a} state. 
\end{abstract}
%
%
\maketitle 
%
%
%
%
%
\section{Introduction}
The process of photoassociation (PA) has served as a very useful tool to researchers in the field of ultracold Atomic, Molecular and Optical(AMO) physics.\cite{Book09} The main advantage of this technique is its simplicity. The PA process provides an efficient, continuous, single-step method for the formation of alkali metal molecules in the ground $X\,^1\Sigma^+$ state and the metastable $a\,^3\Sigma^+$ state.\cite{Wang04a} Detection of the molecules formed by PA and subsequent spontaneous emission is possible using several techniques, including resonance enhanced multiphoton ionization (REMPI). The REMPI process provides very rich spectra containing information on not only the vibrational population distribution of the \emph{X} state and the \emph{a} state but also on the excited intermediate state via which ionization is performed. 

A major focus of the AMO community in recent years has been on the production of the absolute ground state of ultracold heteronuclear molecules. Such molecules may well be a stepping stone for the future realization of quantum computation and improved understanding of many-body physics. Researchers all over the world have searched for different transfer pathways for the formation of these ground-state molecules, e.g. photoassociation \cite{Deiglmayr08,Zabawa11} and stimulated Raman adiabatic passage (STIRAP). \cite{Aikawa10, Ospelkaus10}

Our group at the University of Connecticut has sought an efficient path for formation of absolute ground-state molecules of $^{39}$K$^{85}$Rb using single-step PA. \cite{Banerjee10} Accurate knowledge of the excited electronic states is important to this research since the available \emph{ab initio} potentials are not sufficiently accurate. Hence, we have performed high resolution spectroscopy of various electronic states (both excited and ground) of KRb over the years. \cite{Wang04a,Wang04b,Wang05,Wang06,Wang07,Banerjee12, JTKim09, JTKim11, JTKimPRA11, JTKim12} In this paper, we report both experimental and theoretical investigations of the double-minimum $3\,^3\Pi$ electronic state of $^{39}$K$^{85}$Rb, which may be useful in both the formation and detection of the \emph{X} state and the \emph{a} state. 
\section{Experiment}

\begin{figure}[h]
\centering \vskip 0 mm
\includegraphics[clip, width=\linewidth]{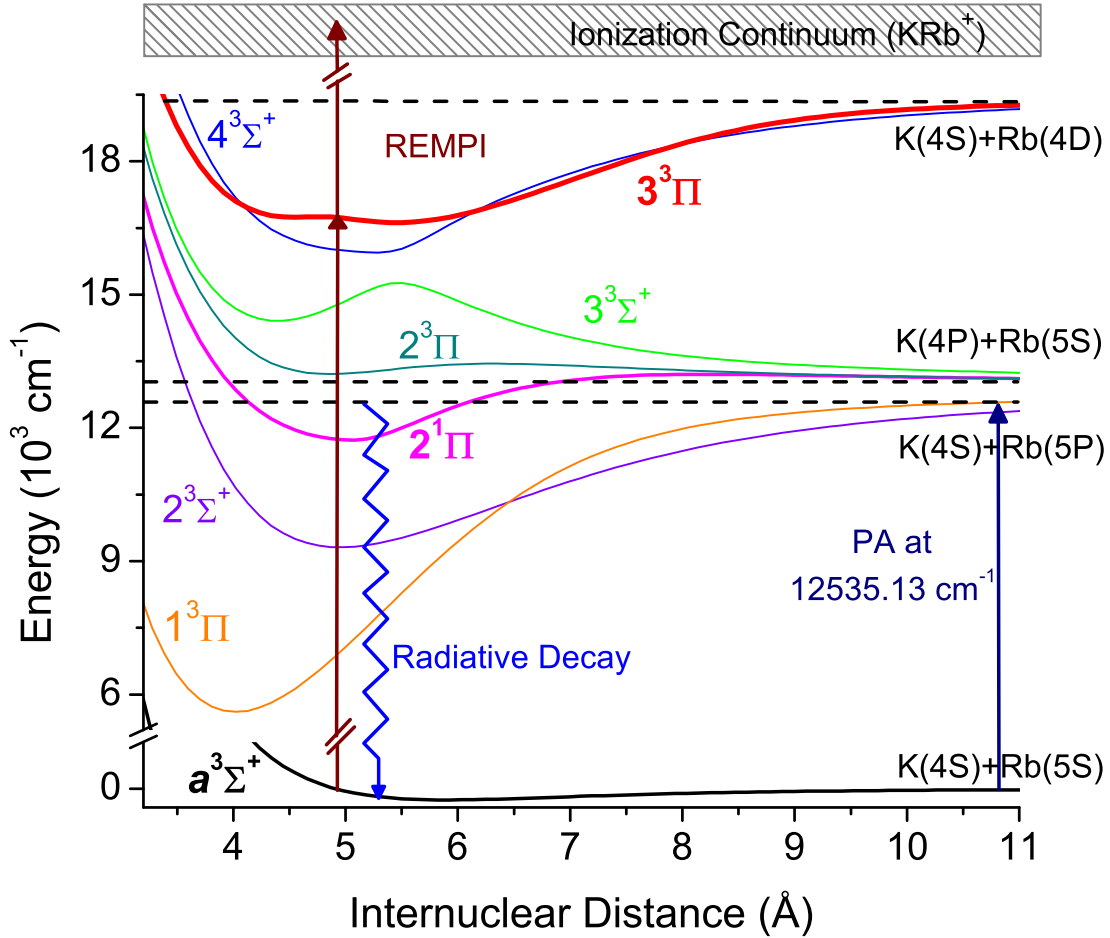}
\caption{\protect\label{Fig1} Scheme for detection of the double-minimum $3\,^3\Pi$ excited electronic state of KRb. The PA laser is detuned below the K$(4s)$ + Rb$(5p_{1/2})$ asymptote by 43.82 cm$^{-1}$. The resulting $a\,^3\Sigma^+$ state molecules are then ionized via REMPI, where the $3\,^3\Pi$ state is the resonant intermediate state.} 
\end{figure}

A detailed description of the experimental setup is given elsewhere. \cite{Wang04b} Here we give a brief account. We start with a dual species $``$Dark-SPOT'' \cite{Ketterle93} magneto-optical trap (MOT) of $^{39}$K and $^{85}$Rb. The average MOT density of $^{39}$K is $\sim$3$\times10^{10}$  cm$^{-3}$ and that of $^{85}$Rb is $\sim$1$\times10^{11}$ cm$^{-3}$. After the two MOTs are carefully overlapped, we introduce a photoassociation laser to form KRb molecules. The PA laser is a cw titanium:Sapphire ring laser pumped by a 10 W 532 nm laser (Coherent Radiation model Verdi V10). The typical output of the PA laser is $\sim$1 W. The PA process produces excited state KRb molecules, which then undergo radiative decay to form molecules in the ground $X\,^{1}\Sigma^+$ state and the metastable $a\,^3\Sigma^+$ state. Finally, the ground \emph{X} and metastable \emph{a} state KRb molecules are ionized using the REMPI method with time-of-flight detection by a Channeltron charged particle detector. The ionization laser is a Continuum model ND6000 pulsed dye laser pumped at 532 nm by a doubled Nd:YAG laser. The REMPI laser has a linewidth of 0.5 cm$^{-1}$ and produces 10 ns pulses of $\sim$1 mJ energy, focused to  a diameter (FWHM) of $\sim$0.76 mm at the MOT. The KRb$^+$ ions thus produced are distinguished from other ions (Rb$^+$, K$^+$, Rb$_2^+$) using time of flight mass spectroscopy.

Electronic and vibrational spectroscopic information on the $3\,^3\Pi_{\Omega}$ states is obtained by scanning the REMPI laser while keeping the PA laser fixed at the desired frequency. The experimental scheme is shown in Figure 1. The PA laser frequency is fixed at 12535.13 cm$^{-1}$, populating the $v'=17, J'=2$ level of the 5(1) state, which at short range corresponds to the $2\,^1\Pi$ state. For analysis of the $3\,^3\Pi_{\Omega}$ potentials we use diabatic potentials obtained by modifying the existing \emph{ab initio} potentials calculated by Rousseau \emph{et al}. \cite{Rousseau00} The lowest triplet-state potential correlating with two ground-state atoms is determined from experimental work. \cite{Pashov07}
\section{Modified \emph{Ab Initio} Potential Energy Curves}

\begin{figure*}[h]
\centering \vskip 0 mm
\includegraphics[clip, width=0.5\linewidth]{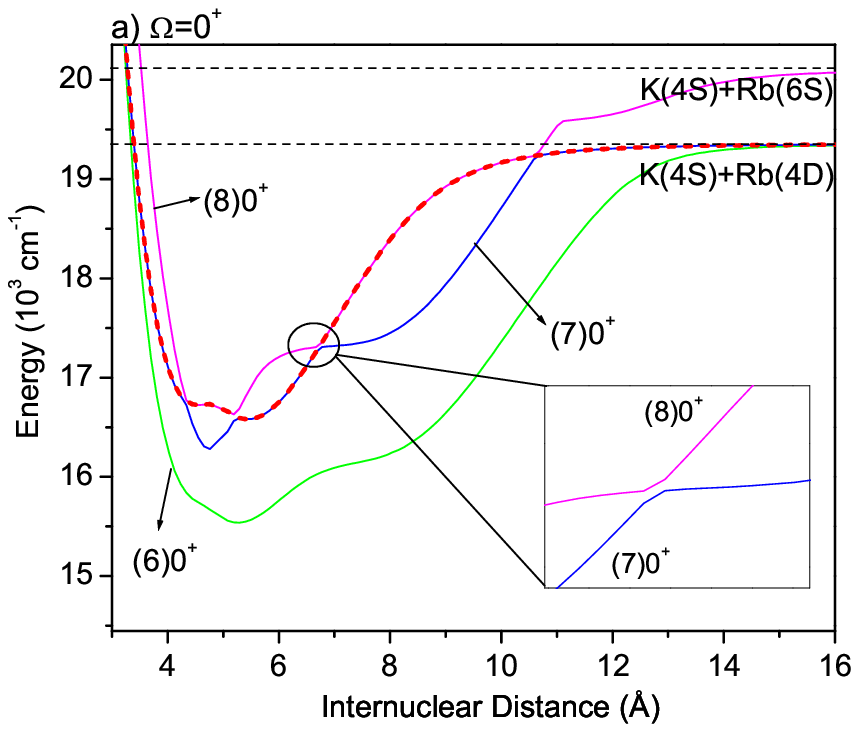}%
\includegraphics[clip, width=0.5\linewidth]{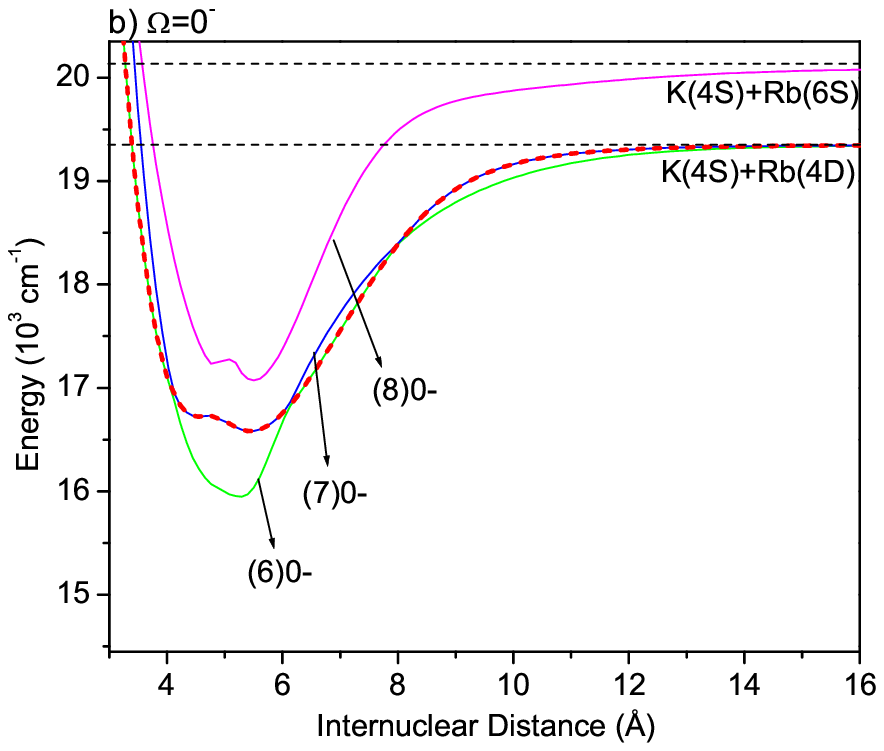}\\
\includegraphics[clip, width=0.5\linewidth]{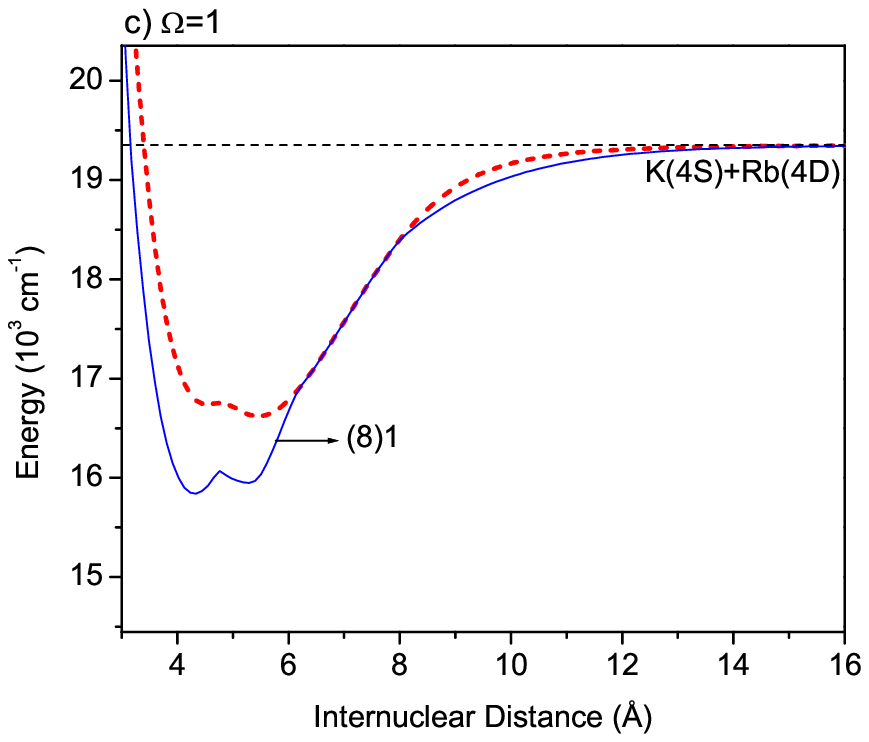}%
\includegraphics[clip, width=0.5\linewidth]{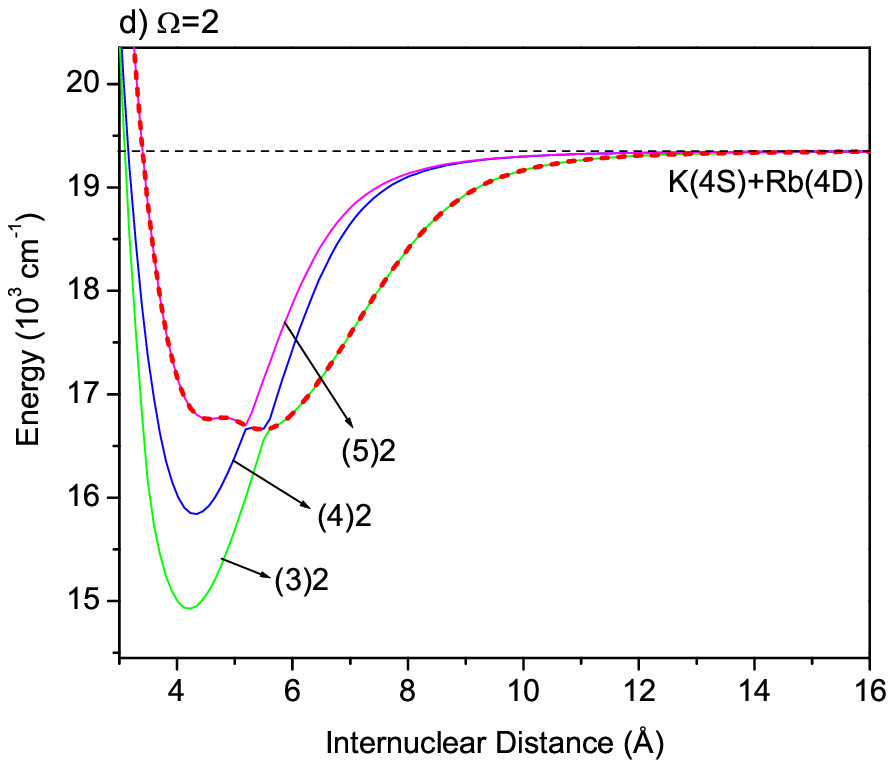}%
\caption{\protect\label{Fig2} Plots of all available theoretical Hund's case (c) potentials arising from the K(4$s$)$+$Rb(4$d$) and K(4$s$)$+$Rb(6$s$) asymptotes . The $\Omega=0^+, 0^-, 1$ and 2 components are shown separately. In each case, the dashed line curve indicates the diabatic potential constructed to approximate the corresponding $\Omega$ component of the $3\,^3\Pi$ state. The inset in (a) shows an expanded view of a typical avoided crossing between adiabatic curves.} 
\end{figure*}

\begin{figure}[h]
\centering \vskip 0 mm
\includegraphics[clip, width=\linewidth]{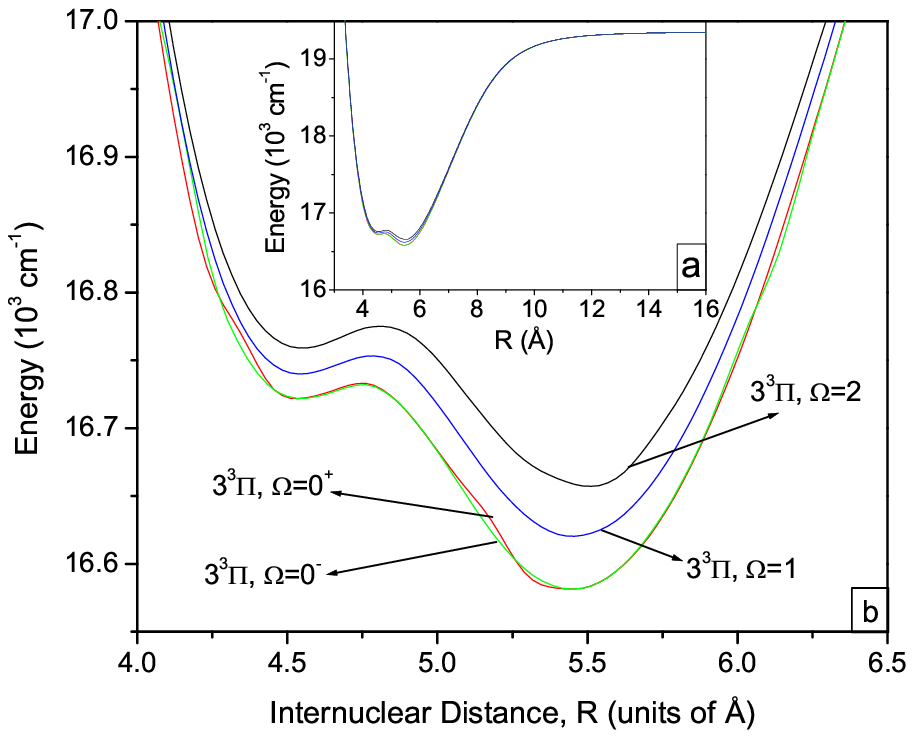}
\caption{\protect\label{Fig3} Modified potential energy curves showing the spin-orbit components of the $3\,^3\Pi$ state. The inset (a) shows the entire potential curves, while the main figure (b) shows an expanded view in which the $\Omega=0,1$ and 2 components are distinct. The $\Omega=0^{\pm}$ components, as expected, are nearly degenerate.} 
\end{figure}

Due to spin-orbit interactions, the $3\,^3\Pi$ state splits into four separate potentials with $\Omega= 0^{+}, 0^{-}, 1$ and 2. We find that the $\Omega=0^+$ and $0^-$ components are split by much less than the vibrational spacing, so it suffices to determine three potential curves for $3\,^3\Pi_\Omega$ with $\Omega = 0$, 1, and 2. Unfortunately, there are no direct \textit{ab initio} calculations of the requisite Hund's case(a) potentials.  Instead, the available \textit{ab initio} potentials include a calculation of the potential at the short-internuclear range notation (appropriate for Hund's case (b)) for the $3\,^3\Pi$ state without spin-orbit terms\cite{Rousseau00}, as well as calculations of pure case (c) potentials that correspond to the $\Omega = 0^\pm$ and 2 components, but not $\Omega=1$. To construct approximate case (a) potentials for the analysis of our experimental data, we modify the existing Hund's case c) \textit{ab initio} potentials where available, and use the case (b) potential to approximate the $\Pi_1$ component.

Figure 2 shows the all of the available case (c) potentials converging to the K(4$s$)$+$Rb(4$d$) and K(4$s$)$+$Rb(6$s$) atomic asymptotes\cite{Rousseau00}, with panels (a), (b), and (d) depicting $\Omega=0^+, 0^-$, and 2. There are numerous avoided crossings, and in the case (a) limit it is the diabatic continuation of each curve across these gaps that is needed.  We have constructed approximate potential curves for the $\Pi_{0^-}, \Pi_{0^+}$, and $\Pi_2$ components of the $3\,^3\Pi$ state by using cubic splines to smoothly connect the adiabatic curves at the crossings.  The dashed curves in Fig. 2 show the resulting case (a) potentials.  For the $\Pi_1$ component this procedure is impossible because the required 9(1) state has not been calculated, so instead we approximate the case (a) potential by directly using the calculated potential curve of the Hund's case (b) $3\,^3\Pi$ state, which is shown by the dashed curve in panel (c).  This approximation is reasonable because the case (b) potential curve represents the weighted average of all spin-orbit components, which should lie close to the middle $\Pi_1$ component. It is also in good agreement with our experimental results, as we show in Section IV.
 
In Figure 3, all four of the resulting approximate potential energy curves are shown in an expanded view, together with an inset showing their entire range. The near degeneracy of the $\Omega=0^\pm$ components is evident, as is typical in Hund's case (a), and from now on we will refer to them together as $\Omega=0$. Our experimental observations confirm that the $\Omega=0^\pm$ splitting is quite small. We have provided the listing of the modified potentials in the supplementary material.\cite{Note1}

\begin{figure}[h]
\centering \vskip 0 mm
\includegraphics[clip, width=\linewidth]{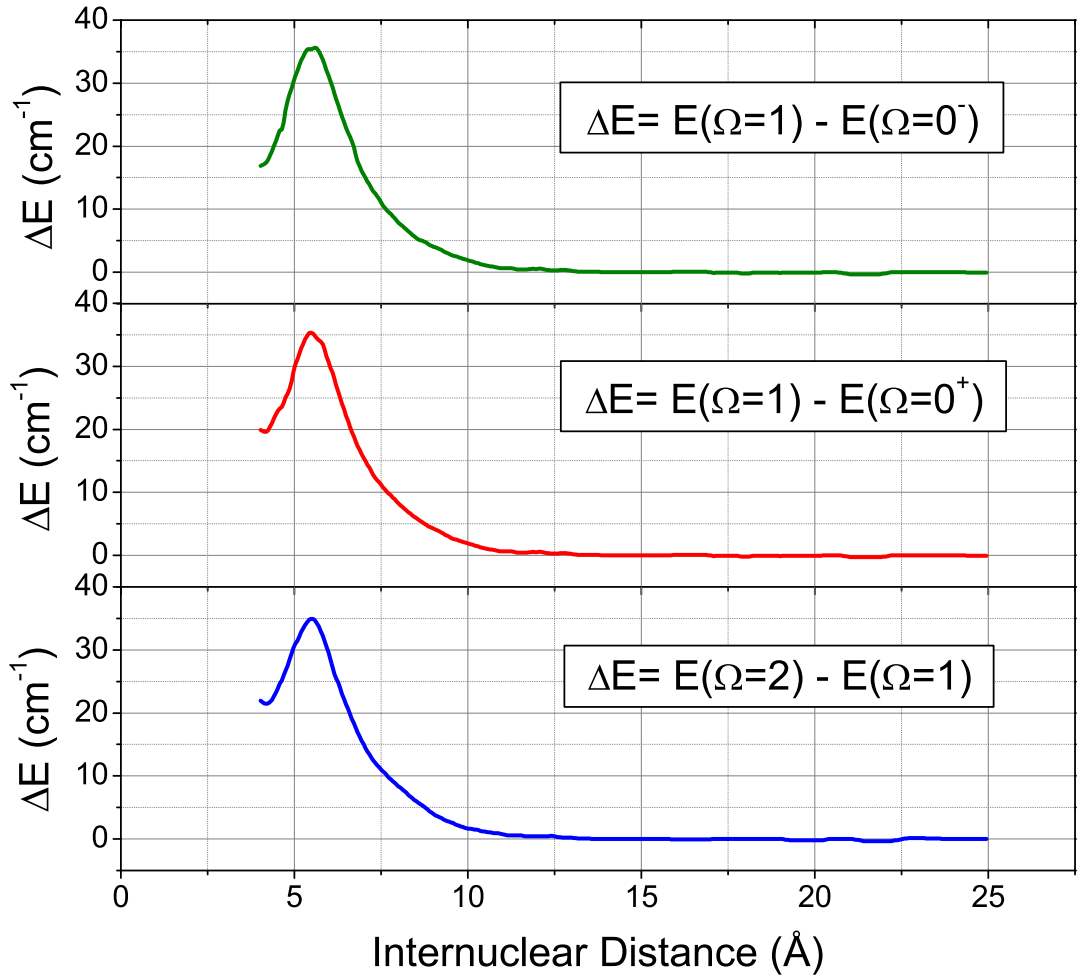}
\caption{\protect\label{Fig4} Variations of the energy splittings between different spin-orbit components of the $3\,^3\Pi$ state with respect to internuclear distance. The splittings are largest near $\sim$5 \AA\;, corresponding to the bottom of the well.} 
\end{figure}

At long range the $3\,^3\Pi$ potentials correlate to the K(4$^2$S$_{1/2}$)+Rb(4$^2$D$_J$) atomic states, and the fine-structure splitting between $J=3/2$ and $J=5/2$ is only 0.44 cm$^{-1}$. The much larger molecular fine-structure must gradually converge to this small value as the potentials approach their asymptotes. This is quantitatively represented in Figure 4, which shows that the spin-orbit splittings are largest ($\sim$40 cm$^{-1}$) near the bottom of the well ($\sim$5 \AA) and decrease rapidly at long range.

\section{Results and Analysis}

In this section, we first describe our PA and REMPI spectra, then analyze them using the potentials described in section III. We also discuss possible perturbations due to vibrational near-degeneracy between levels of the $3\,^3\Pi_{\Omega}$ and the $4\,^3\Sigma^+$ states.

\subsection{Photoassociation Spectra}

\begin{figure}
\centering \vskip 0 mm
\includegraphics[clip, width=\linewidth]{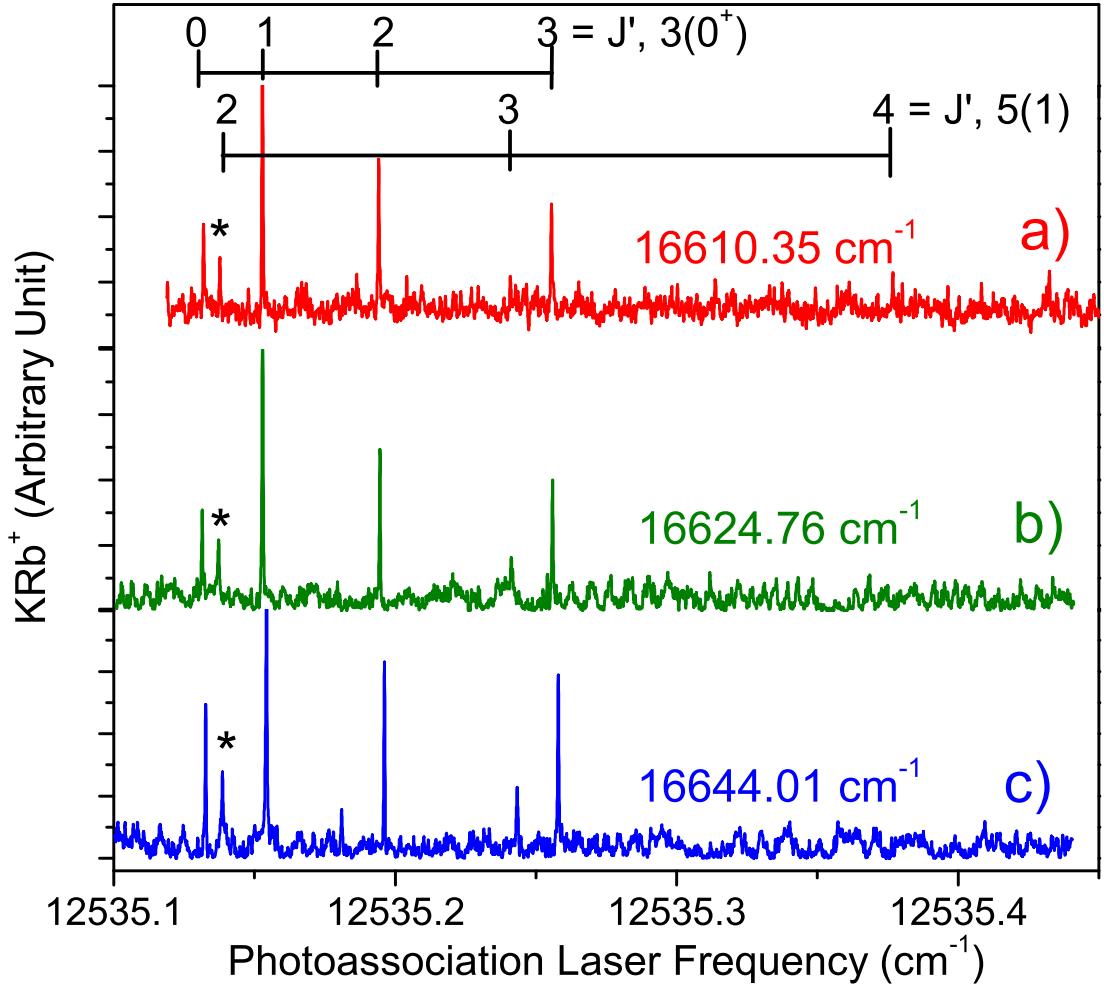}
\caption{\protect\label{Fig5} PA spectra indicating the overlapping $3(0^+)$ and 5(1) bands, with the REMPI detection laser frequency fixed at a) 16610.35 cm$^{-1}$ b) 16624.76 cm$^{-1}$ and c) 16644.01 cm$^{-1}$. The three PA spectra confirm the reproducibility of the spectral features. The 5(1), $J'=2$ line is from $v'=17$ and correlates to the short-range $2\,^1\Pi$ state. \cite{Banerjee10} This is the excited state that produces $a\,^3\Sigma^+$ state molecules with $v''=15-25$, and is marked by an asterisk.} 
\end{figure}

Figure 5 shows PA spectra near 12535.1 cm$^{-1}$. In these spectra two rotational bands are identified. The stronger KRb$^+$ signals correspond to $J'=0-3$ of the $3(0^+)$ state with $v'=118$ (the $1\,^3\Pi$ state at short range) and the weaker to $J'=2-4$ of the 5(1) state with $v'=17$ (the $2\,^1\Pi$ state at short range), for which $J'=2$ and 3 are most prominent. We were not able to observe the $J'=1$ level of the 5(1) state in any of our PA scans. The set of three PA spectra with three different detection laser frequencies confirms the reproducibility of the spectral features both for the strong and the weak signals. The assignments of the PA spectra are based on previous experimental observation of the $3(0^+)$, $v'$ levels by photoassociation in a MOT. \cite{Wang04a} The 5(1), $v'$ levels were experimentally observed for higher values of $J'$ by optical-optical double resonance polarization spectroscopy. \cite{Kasahara99} They were extrapolated to predict the positions of the 5(1), $v', J'=1-4$ levels. \cite{Banerjee10} 

The vibrational numbering of the 5(1) state is unambiguous since numerous levels were previously experimentally observed.\cite{Kasahara99} However, for the $3(0^+)$ state, we have previously observed only five vibrational levels of the $3(0^+)$ state near the asymptote.\cite{Wang04a} so the vibrational numbering is uncertain. Calculating vibrational levels of the $3(0^+)$ state using LEVEL \cite{LEV74} and the \emph{ab initio} potential, we find that $v'=118$ of $3(0^+)$ is the closest match to the present photoassociation line. Thus we will refer to the $3(0^+)$ band reported in Figure 5 as $v'=118$, though the absolute numbering remains uncertain.

In Appendix A we discuss possible problems due to overlaps with $``$hyperfine ghost'' artifacts, arising due to a small population in the bright hyperfine states (F = 2 for $^{39}$K and F = 3 for $^{85}$Rb) of $^{39}$K (4S) and $^{85}$Rb (5S) in our dark SPOT, and we rule out the possibility.

\subsection{REMPI Spectra}

\begin{figure*}[h]
\centering \vskip 0 mm
\includegraphics[clip, width=0.8\linewidth]{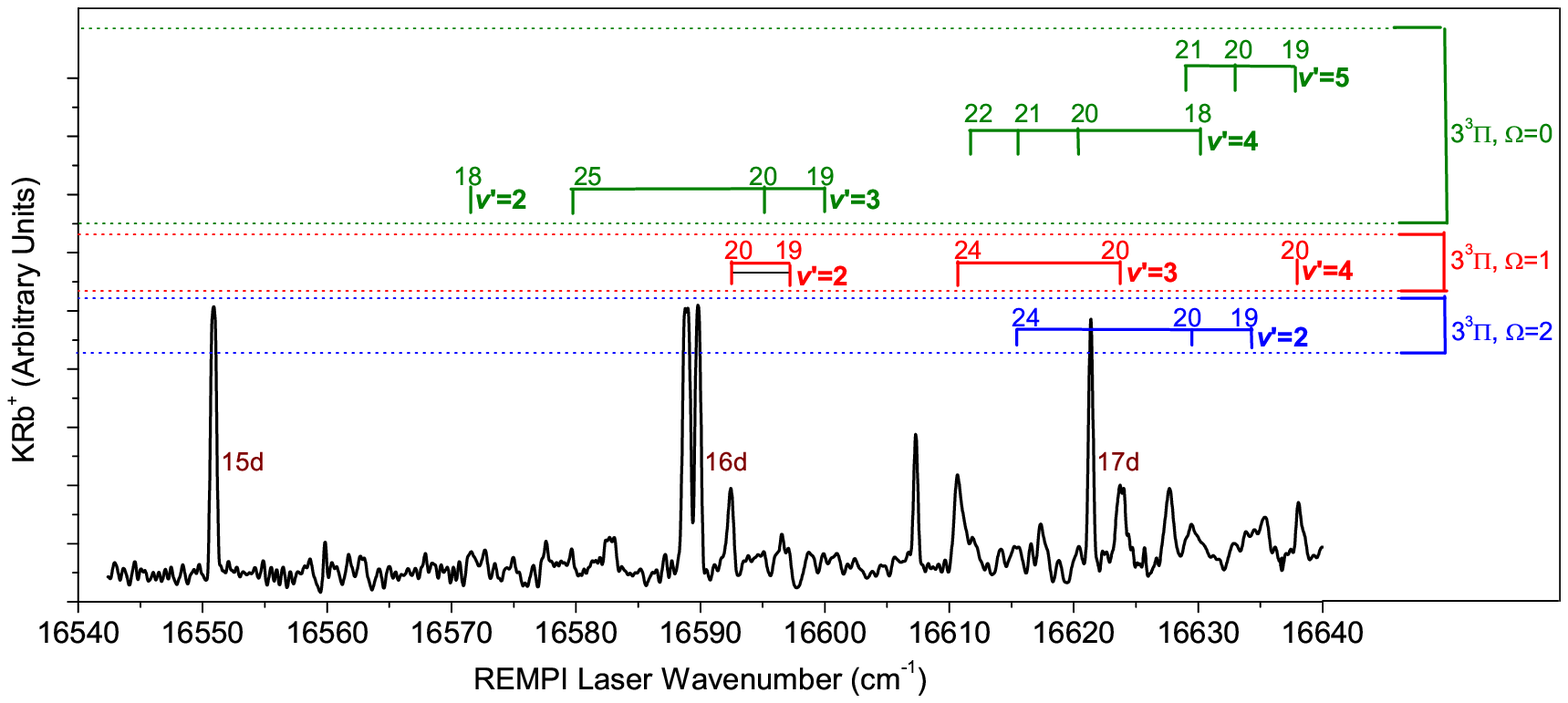}\\
\vspace{-1cm}
\includegraphics[clip, width=0.8\linewidth]{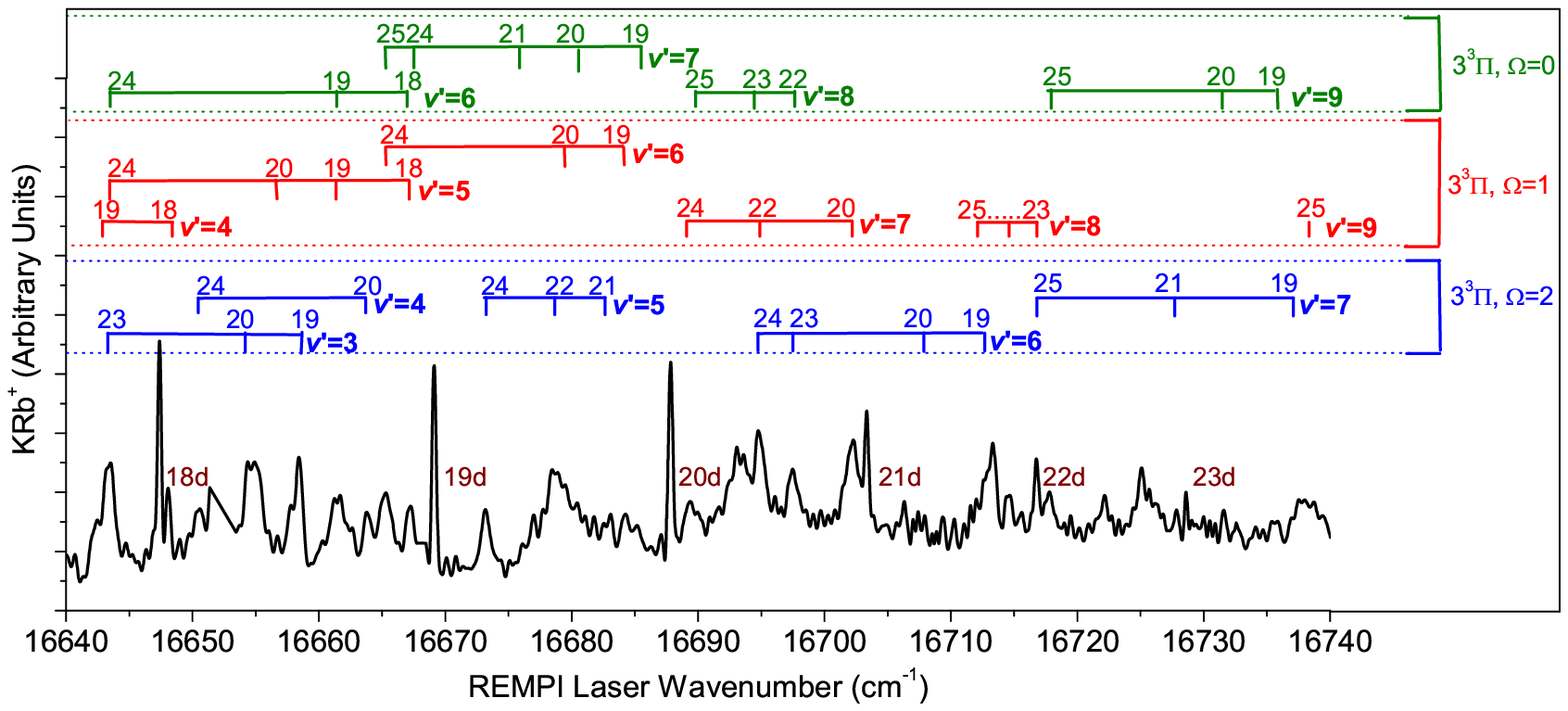}\\
\vspace{-1cm}
\includegraphics[clip, width=0.8\linewidth]{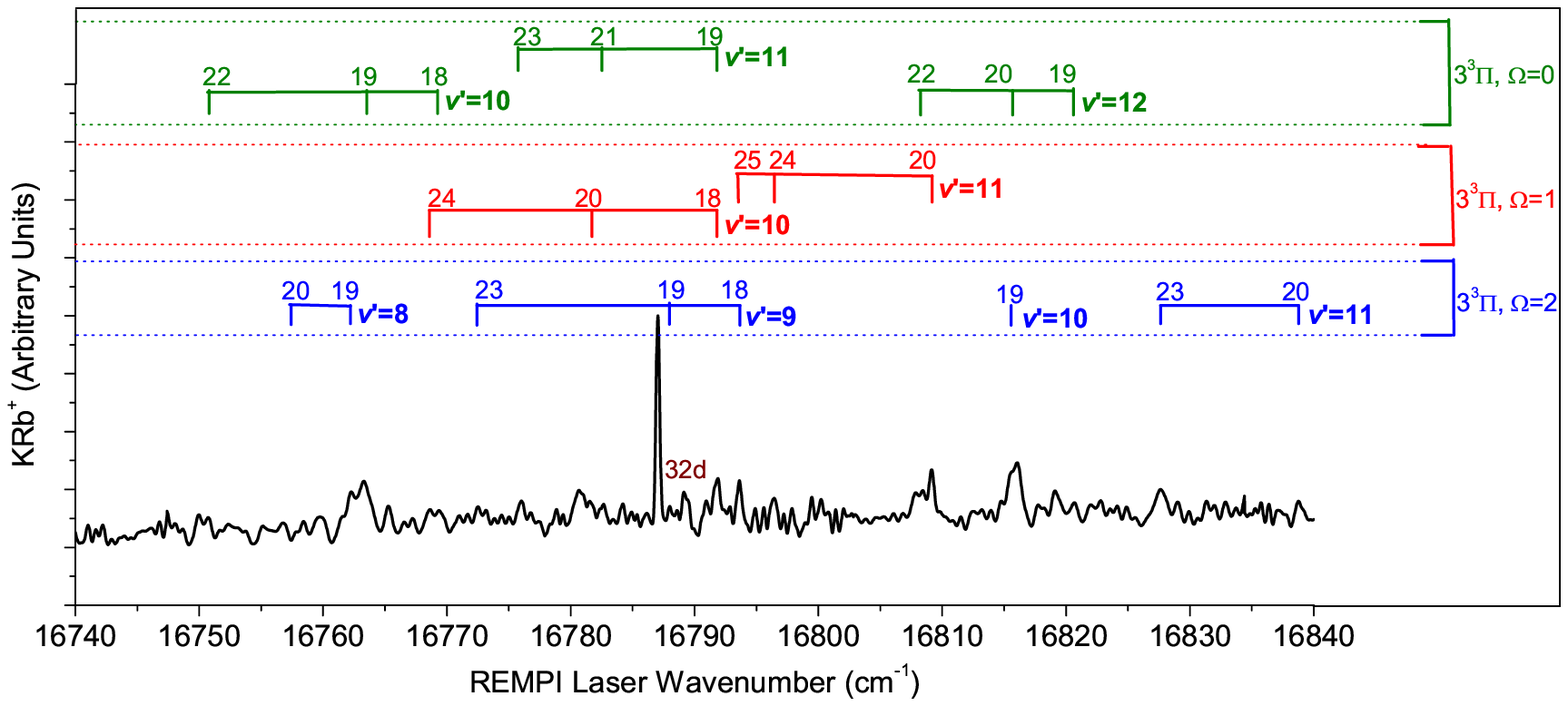}
\caption{\protect\label{Fig6} REMPI spectra indicating spectral features corresponding to  $3\,^3\Pi_{\Omega}$$\leftarrow$$a\,^3\Sigma^+$ 
transitions (vertical lines) and atomic Rb \emph{nd}$\leftarrow$$5s$ 
two-photon transitions. A variety of $v''$ values from 14 to 25 were observed, but for clarity only values from $v''=18-25$ (above the noise level) are shown. The saturated atomic signal close to Rb 16d$\leftarrow$$5s$ corresponds to a $^{39}$K atomic transition. The PA laser is fixed on the 5(1), $v'=17, J'=2$ level, at 12535.13 cm$^{-1}$.} 
\end{figure*}

The REMPI spectra obtained by keeping the PA laser fixed at the 5(1), $v=17, J=2$ level are shown in Figure 6. The spectra consist of transitions from the $a\,^3\Sigma^{+}, v''=15-25$ levels to the $3\,^3\Pi_{0}, v'=2-12$ levels, the $3\,^3\Pi_{1}, v'=2-11$ levels and the $3\,^3\Pi_{2}, v'=2-11$ levels. The strongest transitions to the $\Omega=1$ and 2 components of the $3\,^3\Pi$ state originate from $v''=20$ and 24 of the $a\,^3\Sigma^{+}$ state, and the strongest transitions to the $\Omega=0$ component of the $3\,^3\Pi$ state originate from $v''=19$ and 20. In Figure 6, we indicate only the transitions which are unambiguously assignable from the vibrational levels $v''=18-25$ of the triplet metastable state to the spin-orbit components of the $3\,^3\Pi$ state. The spectra are calibrated using atomic two-photon resonances, labeled by $``nd$'', which are visible because the long-time tail of the atomic Rb$^+$ time-of-flight distribution leaks into the time gate used for KRb$^+$. The accuracy of the  vibrational energies is determined by the REMPI laser linewidth, which is 0.5 cm$^{-1}$. The REMPI signal strength is dependent on the combination of the efficiency of \emph{a}-state molecule formation in various $v''$ levels by PA via the 5(1), $v=17, J=2$ level and the efficiency of the bound-bound transitions between the vibrational levels of the $a\,^3\Sigma^{+}$ and the $3\,^3\Pi_{\Omega}$ states.

\subsection{Analysis and Discussion}

\begin{table}[h]
\caption{\label{tab:Values}Contributions of various $J'$s (of $3\,^3\Pi_{\Omega}$ states)  to the overall lineshape of the REMPI signals.}
\begin{ruledtabular}
\scalebox{1.0}{
\begin{tabular}{cccc}
\multicolumn{1}{c}{$J'$}&\multicolumn{1}{c}{$3\,^3\Pi_0$}&\multicolumn{1}{c}{$3\,^3\Pi_1$}&\multicolumn{1}{c}{$3\,^3\Pi_2$}\\
\hline
0&10\%&0&0\\
1&20\%&20\%&0\\
2&28.58\%&40.48\%&38.1\%\\
3&30\%&21.67\%&33.35\%\\
4&11.43\%&17.8\%&28.57\%\\
\end{tabular}}
\end{ruledtabular}
\end{table}

\begin{figure*}[h]
\centering \vskip 0 mm
\includegraphics[clip, width=0.5\linewidth]{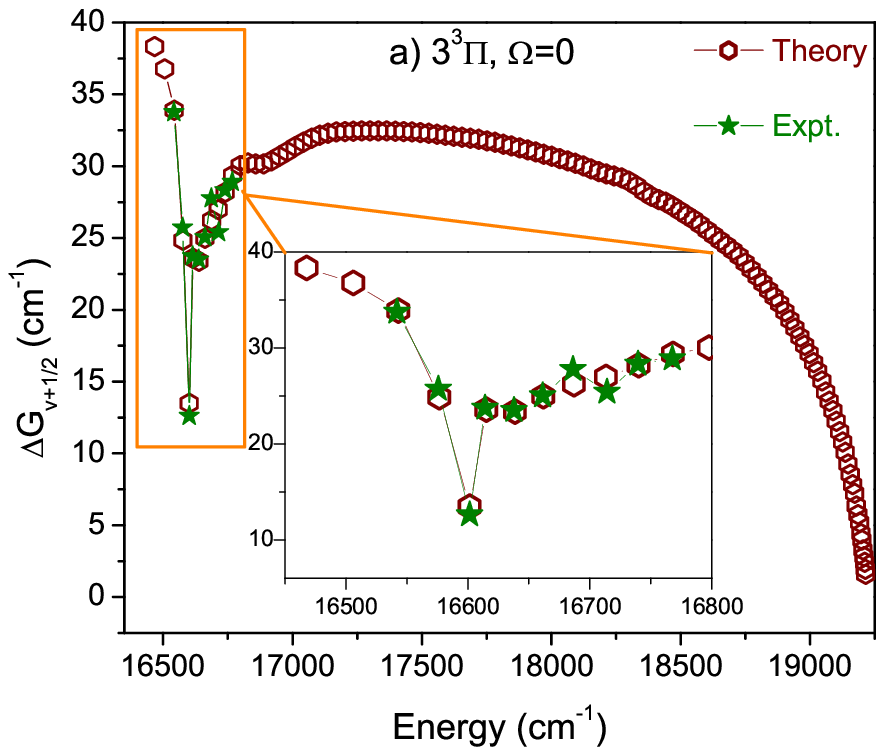}%
\includegraphics[clip, width=0.5\linewidth]{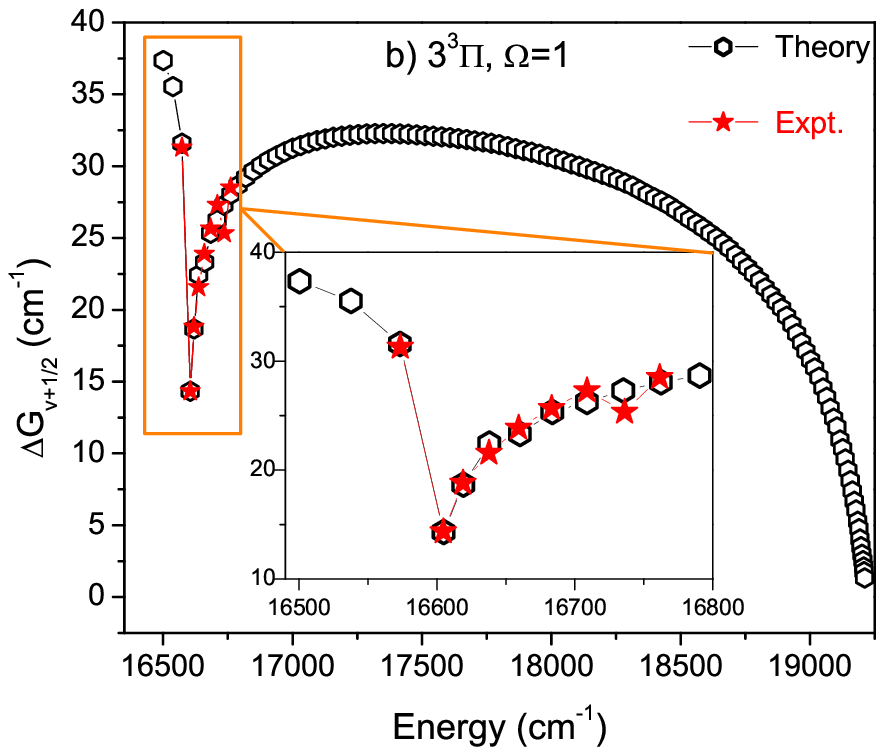}\\
\includegraphics[clip, width=0.5\linewidth]{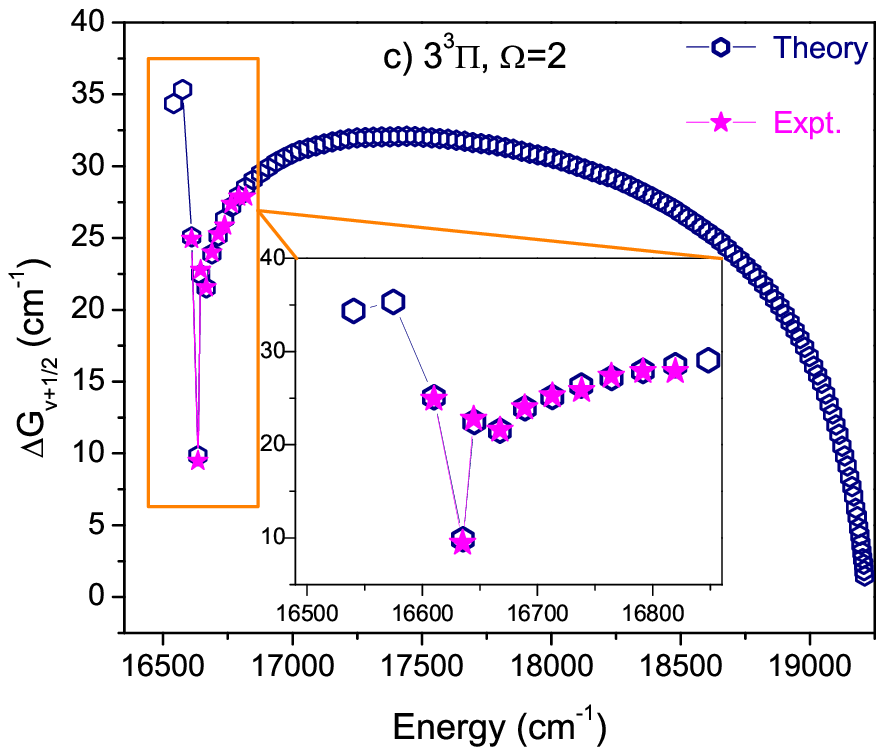}
\caption{\protect\label{Fig7} Comparison of theoretical and experimental values of the vibrational spacings, $\Delta$G$_{v+1/2}$ (cm$^{-1}$) for the levels of $3\,^3\Pi_{\Omega}$ states, where (a) shows the $3\,^3\Pi_{0}$ state, (b) the $3\,^3\Pi_{1}$ state, and (c) the $3\,^3\Pi_{2}$ state. The theoretical values are obtained using LEVEL \cite{LEV74} with the modified potentials reported in section III. The experimental values are obtained from the present work. The insets show an expanded view of the regions indicated by the rectangles (orange).} 
\end{figure*}

We have successfully assigned the vibrational levels $v'=2-11$ of the $3\,^3\Pi_1$ and $3\,^3\Pi_2$ state and $v'=2-12$ of the $3\,^3\Pi_0$ state. From these observed transitions we obtain the total energy (E$_{v'J'}$) for a particular vibrational level $v'$ which has a spread over its rotational levels ($J'$) (these are not resolved due to the laser linewidth). The total energy is given by E$_{v'J'} =$ T$_{v'} + J'(J'+1)$B$_{v'}$ where T$_v'$ is the vibrational term energy and B$_{v'}$ is the rotational constant. Hence to obtain T$_v'$ from E$_{v'J'}$ it is required to estimate the contributions of various $J'$s to the overall lineshape of the transitions. In Table I, for each $\Omega$ component of the $3\,^3\Pi$ state, we estimated contributions of various $J$s, using H\"{o}nl-London factors.\cite{Herzberg} To obtain the numbers in Table I, we first calculated the spread of $J'$s in the \emph{a}-state resulting from the PA process and then the spread of $J'$s in the $3\,^3\Pi$ state due to the REMPI. From the numbers in Table I, it is possible to calculate the approximate shifts in E$_{v'J'}$ values to obtain the T$_v'$ values. For the $3\,^3\Pi_{0}$ state the shift is $\sim$8B$_{v'}$, for the $3\,^3\Pi_{1}$ state the shift is $\sim$8.99B$_{v'}$ and for the $3\,^3\Pi_{2}$ state the shift is $\sim$12B$_{v'}$. For example, the theoretical B$_{v'}$ values (calculated using LEVEL \cite{LEV74}) for $v'=3$ of $3\,^3\Pi_{0}$, $3\,^3\Pi_{1}$ and $3\,^3\Pi_{2}$ states are 0.02235 cm$^{-1}$, 0.02309 cm$^{-1}$ and 0.02652 cm$^{-1}$ respectively. Then the required shift in E$_{v'=3,J'}$ to obtain T$_{v'=3}$ for the $3\,^3\Pi_{0}$, $3\,^3\Pi_{1}$ and $3\,^3\Pi_{2}$ states are 0.18 cm$^{-1}$, 0.21 cm$^{-1}$ and 0.32 cm$^{-1}$ respectively. For our experiment this shift is well within the laser linewdith, so there is no significant difference between E$_{v'J'}$ and T$_v'$.

\begin{table}[h]
\caption{\label{tab:TableI}Comparison of experimental and theoretical term energies $T_{v'}$ values of the vibrational levels $v'=0-12$ of the $3\,^3\Pi_{0^{-}}$ state. We also compare the theoretical vibrational term energies of the $3\,^3\Pi_{0^{-}}$ and $3\,^3\Pi_{0^{+}}$ state, for the vibrational levels $v'=0-12$. The term energies are reported with respect to the K(4S$_{1/2}$)+Rb(5S$_{1/2}$) asymptote and are in units of cm$^{-1}$.}
\begin{ruledtabular}
\begin{tabular}{ccccccc}
\multicolumn{1}{c}{$v'$}&\multicolumn{2}{c}{Experiment}&\multicolumn{2}{c}{Theory}&\multicolumn{1}{c}{$\Delta T_{v'}$\footnote{$\Delta T_{v'} = T_{v'}(0^-) - T_{v'}(0^+)$ [both theoretical values]}}&\multicolumn{1}{c}{$\Delta T_{v'}$\footnote{$\Delta T_{v'} = T_{v'}$(Theory)$ - T_{v'}$(Expt.)}}\\

&$T_{v'}$&$\Delta$G$_{v+1/2}$&$T_{v'}$&$\Delta$G$_{v+1/2}$&&\\ \hline

0&-&-&16600.75&38.32&0.49&-\\
1&-&-&16639.07&36.77&-0.56&-\\
2&16541.96&33.75&16675.83&33.89&-1.56&133.88\\
3&16575.70&25.73&16709.73&24.85&-0.07&134.02\\
4&16601.44&12.62&16734.58&13.48&-0.60&133.15\\
5&16614.06&23.74&16748.07&23.59&-0.46&134.01\\
6&16637.79&23.49&16771.66&23.38&-0.62&133.86\\
7&16661.28&25.04&16795.04&24.98&-1.02&133.75\\
8&16686.33&27.78&16820.01&26.22&-0.85&133.68\\
9&16714.11&25.40&16846.23&26.99&-0.69&132.13\\
10&16739.51&28.23&16873.22&28.18&-0.28&133.71\\
11&16767.74&29.31&16901.40&29.36&0.10&133.66\\
12&16797.05&-&16930.76&30.04&0.66&133.71\\
\end{tabular}
\end{ruledtabular}
\end{table}

\begin{table}[h]
\caption{\label{tab:Values}Comparison of experimental and theoretical term energies $T_{v'}$ of the vibrational levels $v'=0-11$ of the $3\,^3\Pi_1$ state. The term energies are reported with respect to the K(4S$_{1/2}$)+Rb(5S$_{1/2}$) asymptote and are in units of cm$^{-1}$.}
\begin{ruledtabular}
\scalebox{1.0}{
\begin{tabular}{cccccc}
\multicolumn{1}{c}{$v'$}&\multicolumn{2}{c}{Experiment}&\multicolumn{2}{c}{Theory}&\multicolumn{1}{c}{$\Delta T_{v'}$}\\
&$T_{v'}$&$\Delta$G$_{v+1/2}$&$T_{v'}$&$\Delta$G$_{v+1/2}$& Theory-Expt.\\ \hline
0&-&-&16639.40&37.35&-\\
1&-&-&16676.75&35.54&-\\
2&16573.33&31.30&16712.29&31.60&138.96\\
3&16604.63&14.33&16743.89&14.32&139.26\\
4&16618.96&18.52&16758.21&18.67&139.25\\
5&16637.48&22.91&16776.88&22.44&139.40\\
6&16660.39&22.82&16799.32&23.30&138.93\\
7&16683.21&25.66&16822.62&25.33&139.41\\
8&16708.87&27.31&16847.95&26.26&139.08\\
9&16736.18&25.33&16874.21&27.29&138.03\\
10&16761.51&28.52&16901.50&28.05&139.99\\
11&16790.03&-&16929.55&28.68&139.52\\
\end{tabular}}
\end{ruledtabular}
\end{table}

\begin{table}[h]
\caption{\label{tab:Values}Comparison of experimental and \emph{ab initio} term energies $T_{v'}$ of the vibrational levels $v'=0-11$ of the $3\,^3\Pi_2$ double minimum state. The term energies are reported with respect to the K(4S$_{1/2}$)+Rb(5S$_{1/2}$) asymptote and are in units of cm$^{-1}$.}
\begin{ruledtabular}
\scalebox{1.0}{
\begin{tabular}{cccccc}
\multicolumn{1}{c}{$v'$}&\multicolumn{2}{c}{Experiment}&\multicolumn{2}{c}{Theory}&\multicolumn{1}{c}{$\Delta T_{v'}$}\\
&$T_{v'}$&$\Delta$G$_{v+1/2}$&$T_{v'}$&$\Delta$G$_{v+1/2}$& Theory-Expt.\\ \hline
0&-&-&16678.48&34.36&-\\
1&-&-&16712.85&35.33&-\\
2&16610.26&24.93&16748.18&25.08&137.92\\
3&16635.18&9.47&16773.26&9.85&138.08\\
4&16644.65&22.78&16783.11&22.45&138.46\\
5&16667.43&21.47&16805.55&21.53&138.12\\
6&16688.90&24.18&16827.08&23.90&138.18\\
7&16713.08&25.19&16850.98&25.14&137.90\\
8&16738.27&25.73&16876.12&26.34&137.85\\
9&16764.00&27.51&16902.46&27.18&138.46\\
10&16791.51&27.89&16929.63&27.93&138.12\\
11&16819.41&-&16957.56&28.54&138.16\\
\end{tabular}}
\end{ruledtabular}
\end{table}

\begin{figure}[h]
\centering\vskip 0mm
\includegraphics[clip, width=\linewidth]{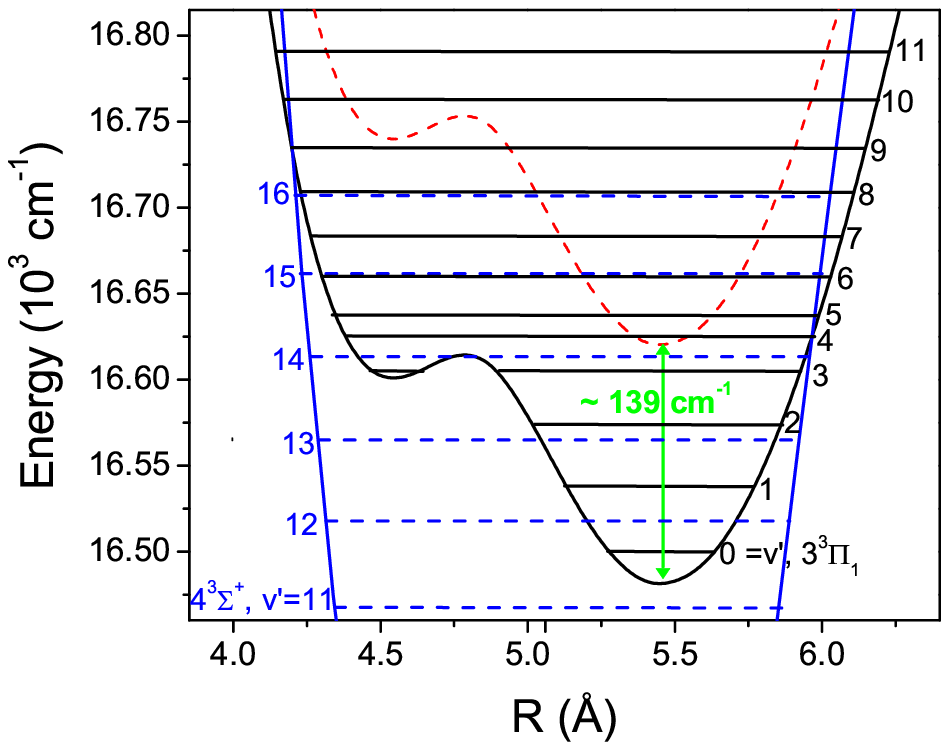}
\caption{\protect\label{Fig8} The solid (black) double minimum potential curve shows the $3\,^3\Pi_1$ state, shifted down by 139 cm$^{-1}$ (indicated by the double headed arrow). The original approximate potential curve is indicated by the dashed (red) line. The solid horizontal lines indicate the vibrational level positions of the $3\,^3\Pi_1$ state for which $v'=0$ and 1 levels are calculated and not experimentally observed. The other solid (blue) potential curve is the $4\,^3\Sigma^+$ state. The horizontal dashed lines indicate the vibrational levels of the $4\,^3\Sigma^+$ state, that have been experimentally observed \cite{Wang06}. Note the near-degeneracy of the $v'=6$ and 8 levels of the $3\,^3\Pi_1$ state with the $v'=15$ and 16 levels of the $4\,^3\Sigma^+$ state, respectively.}   
\end{figure}

The vibrational term energies ($T_{v'}$)  and the vibrational spacings ($\Delta$G$_{v+1/2}$) of each of the $\Omega$ components of the $3\,^3\Pi$ states are tabulated in Tables II, III and IV. In these tables we also compare the experimental values of $T_{v'}$ and $\Delta$G$_{v+1/2}$ with theoretical predictions (calculated using LEVEL \cite{LEV74}) based on the approximate curves described in section III. The calculated values of $T_{v'}$, are higher than the experimental values obtained from experimental data by an average of $\sim$133.6 cm$^{-1}$ for the $3\,^3\Pi_{0}$ state, $\sim$139.3 cm$^{-1}$ for the $3\,^3\Pi_{1}$ state and $\sim$138.1 cm$^{-1}$ for the $3\,^3\Pi_{2}$ state. However the experimental vibrational spacings, $\Delta$G$_{v+1/2}$, are within $6-8\%$ of the theoretical values, as can be seen in Figure 7. The dip in the plot of  $\Delta$G$_{v+1/2}$ vs. energy near 16600 cm$^{-1}$ (Figure 7) is due to the double-minimum character of the potential, evident in Figure 3.

We conclude that the theoretical potentials reported in section III when shifted down by 134-139 cm$^{-1}$, provide an accurate potential energy curve for the double minimum $3\,^3\Pi_{\Omega}$ state in the region of the observed vibrational levels. In Figure 8 we show the $3\,^3\Pi_{1}$ state, where the dashed curve indicates the \emph{ab initio} potential and the solid curve shows this potential shifted down by 139 cm$^{-1}$. 

Also shown in Figure 8 is the portion of the $4\,^3\Sigma^+$ potential curve that overlaps with the $3\,^3\Pi_{\Omega=1}$ state. The $4\,^3\Sigma^+, v'=1-16$ levels were previously experimentally observed \cite{Wang06}. As can be seen in the figure, some vibrational levels of these two states lie in close proximity. Pairs closer than $\sim$2 cm$^{-1}$ include the $4\,^3\Sigma^+, v'=15 - 3\,^3\Pi, v'=6$ and the $4\,^3\Sigma^+, v'=16 - 3\,^3\Pi, v'=8$. This suggests that these levels may appreciably perturb each other. Perturbations may also be significant for the unobserved $v'=17$ level of the $4\,^3\Sigma^+$ state and the $v'=10$ level of the $3\,^3\Pi$ state. However, these perturbations are not obvious in the data in Table III given the $\sim$0.5 cm$^{-1}$ uncertainty from the pulsed laser linewidth. Higher resolution studies would be desirable.

\begin{figure}[h]
\centering\vskip 0mm
\includegraphics[clip, width=\linewidth]{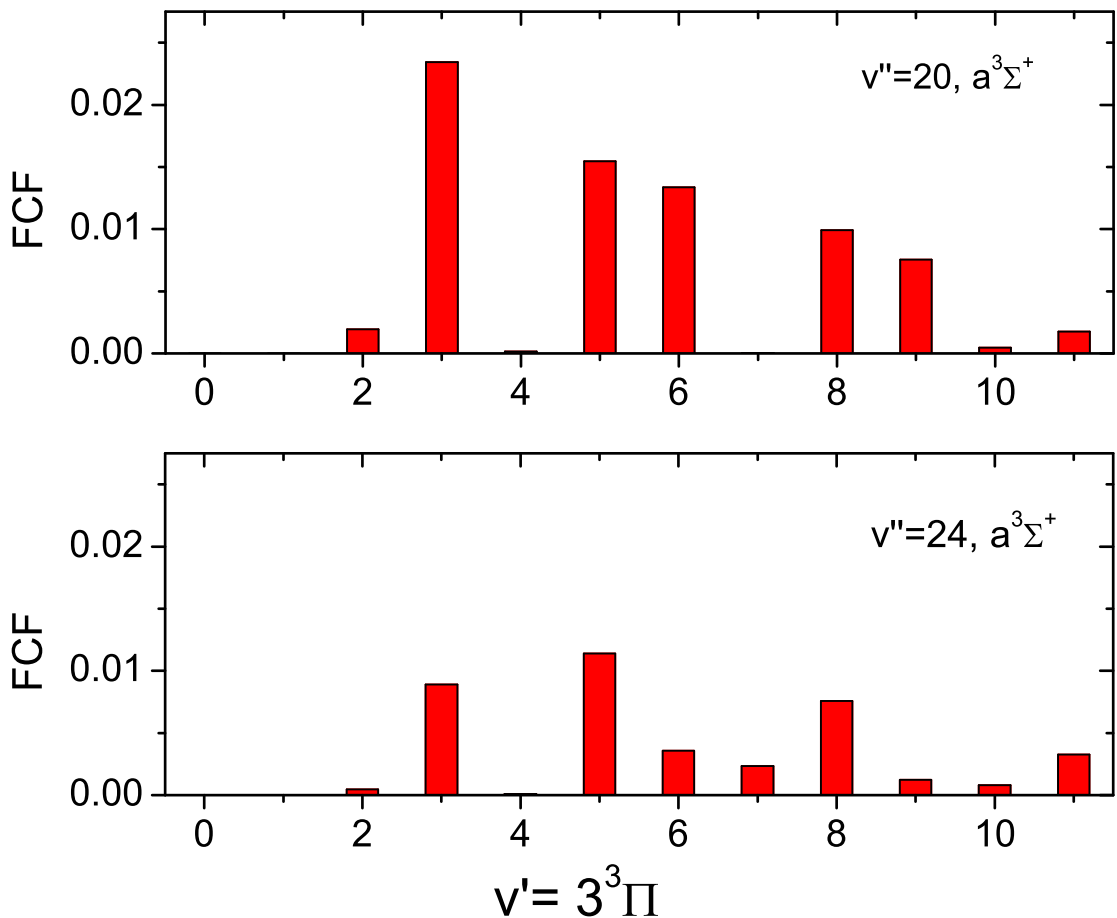}
\caption{\protect\label{Fig9} The FCFs for $3\,^3\Pi_{1}, v'=0-11$$\leftarrow$$a\,^3\Sigma^+, v''=20$ 
and 24, calculated using LEVEL \cite{LEV74}.}   
\end{figure}

We have not observed levels above $v'=12$ of the $3\,^3\Pi_{\Omega}$ states, only because REMPI scans were not performed beyond the reported range. Our inability to observe the levels $v'=0$ and 1 is probably due to the lack of Franck-Condon overlap of these two levels with the $a\,^3\Sigma^+, v''$ levels. This is evident if we focus on the inner turning points (R$_{v-}$) of the $a\,^3\Sigma^+$ and $3\,^3\Pi$ states. As previously discussed, the $a\,^3\Sigma^+$ state is mainly populated in the range $v''=15-25$. The corresponding inner turning points are R$_{15''-}$=4.99 \AA\; and R$_{25''-}$=4.92 \AA. However Figure 8 indicates that the $3\,^3\Pi_{1}, v'=0$ and 1 levels lie deep in the outer well at a distance larger than 4.99 \AA, yielding a poor Franck-Condon overlap. These Franck-Condon Factors (FCFs) have been calculated using LEVEL \cite{LEV74} and are shown in Figure 9 for transitions from $a\,^3\Sigma^+, v''=20$ and 24 to $3\,^3\Pi_{1}, v'=0-11$. The FCFs for the $3\,^3\Pi_{1}, v'=0$ and 1 levels are negligible and for the $v'=2$ level they are relatively weak, in agreement with our experimental observations and the above argument.

\begin{figure}[h]
\centering\vskip 0mm
\includegraphics[clip, width=\linewidth]{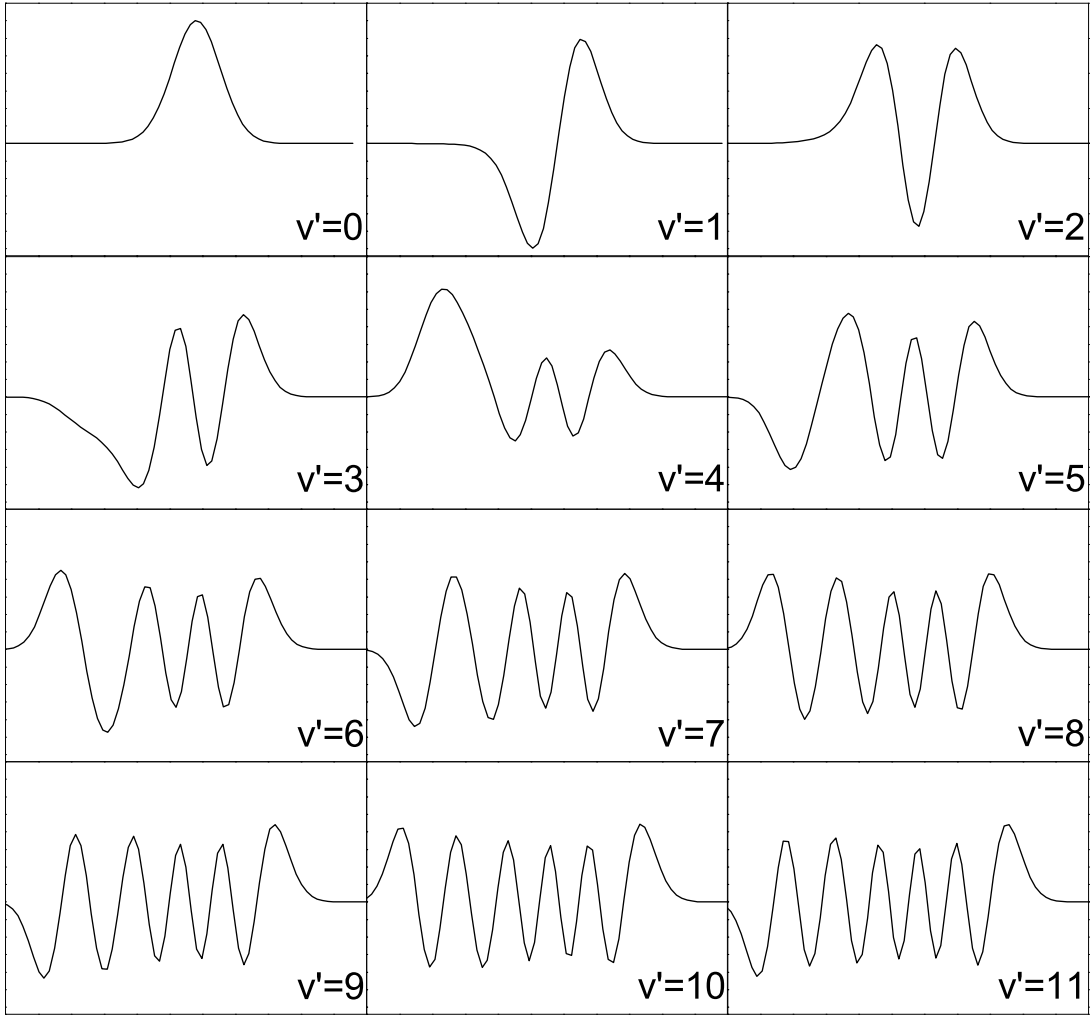}
\caption{\protect\label{Fig10} The vibrational wavefunctions of the $3\,^3\Pi_{1}$ double-minimum state for $v'=0-11$, calculated using LEVEL \cite{LEV74}.}   
\end{figure}

Also, to help characterize this double-well potential, Figure 10 shows the vibrational wavefunctions of this state for all the observed vibrational levels. The outer well is deeper than the inner well as shown in Figure 8. Thus the $v'=0-2$ levels belong entirely to the outer well, while the $v'=3$ wavefunction has a small penetration into the inner well, which is apparent from both its level position (Figure 7) and wavefunction (Figure 10). Above $v'=3$, the amplitude of the wavefunction gradually increases in the inner well region and from $v'=7$ onwards the wavefunctions have a smooth envelope spanning both wells. 

The other spin-orbit components ($\Omega=0$ and 2) of the $3\,^3\Pi$ state should follow very similar arguments to those just  discussed for $\Omega=1$. 
\section{Conclusion}
In conclusion, we have used REMPI of ultracold $^{39}$K$^{85}$Rb to observe the double minimum $3\,^3\Pi_{\Omega}$ states vibrational levels $v'=2-11$ of the $\Omega=1$ and 2 components and $v'=2-12$ of the $\Omega=0$ spin component. Molecules in the $a\,^3\Sigma^+$ state are formed by use of a PA laser fixed at 12535.13 cm$^{-1}$ which corresponds to the $5(1), v'=17, J'=2$ level. We are able to assign the spin-orbit components of the $3\,^3\Pi$ state based on theoretical potentials obtained by modifying the existing \emph{ab initio} potentials. \cite{Rousseau00} Also, we predict that the $4\,^3\Sigma^+$ and the $3\,^3\Pi_{\Omega}$ states may perturb each other at certain vibrational levels which needs further investigation. 

This new spectroscopic information for the excited electronic states of KRb provides better understanding of this molecule. It removes the ambiguity in the electronic and vibrational energies of the excited states in this energy region, thus providing new opportunities for different transfer pathways for future  experiments on ultracold molecules. 
%
%
%
\begin{acknowledgments}
We gratefully acknowledge support from the National Science Foundation and the Air Force Office of Scientific Research (MURI).
\end{acknowledgments}
\appendix
\section{}
The 5(1), $v'=17$ rotational band lies very close to the predicted positions of $^{39}$K hyperfine ghosts of $3(0^+), v'=118$. Here, we provide further analysis to confirm our assignments of the 5(1), $v'=17, J'=2-4$ levels. 

Figure 11 is the PA spectra shown in Figure 5b with additional labels showing the predicted positions for $^{39}$K hyperfine ghosts of $3(0^+), v'=118$. As can be seen, the 5(1), $J'=2$ level lies just below the hyperfine ghost position for $3(0^+), J'=1$, while the 5(1), $J'=3$ is just above the hyperfine ghost position for $3(0^+), J'=3$. Furthermore, there is no clear evidence of any hyperfine ghosts for $3(0^+), J'=0,2$, where no near overlaps occur. 

\begin{figure}[h]
\centering\vskip 0mm
\includegraphics[clip, width=\linewidth]{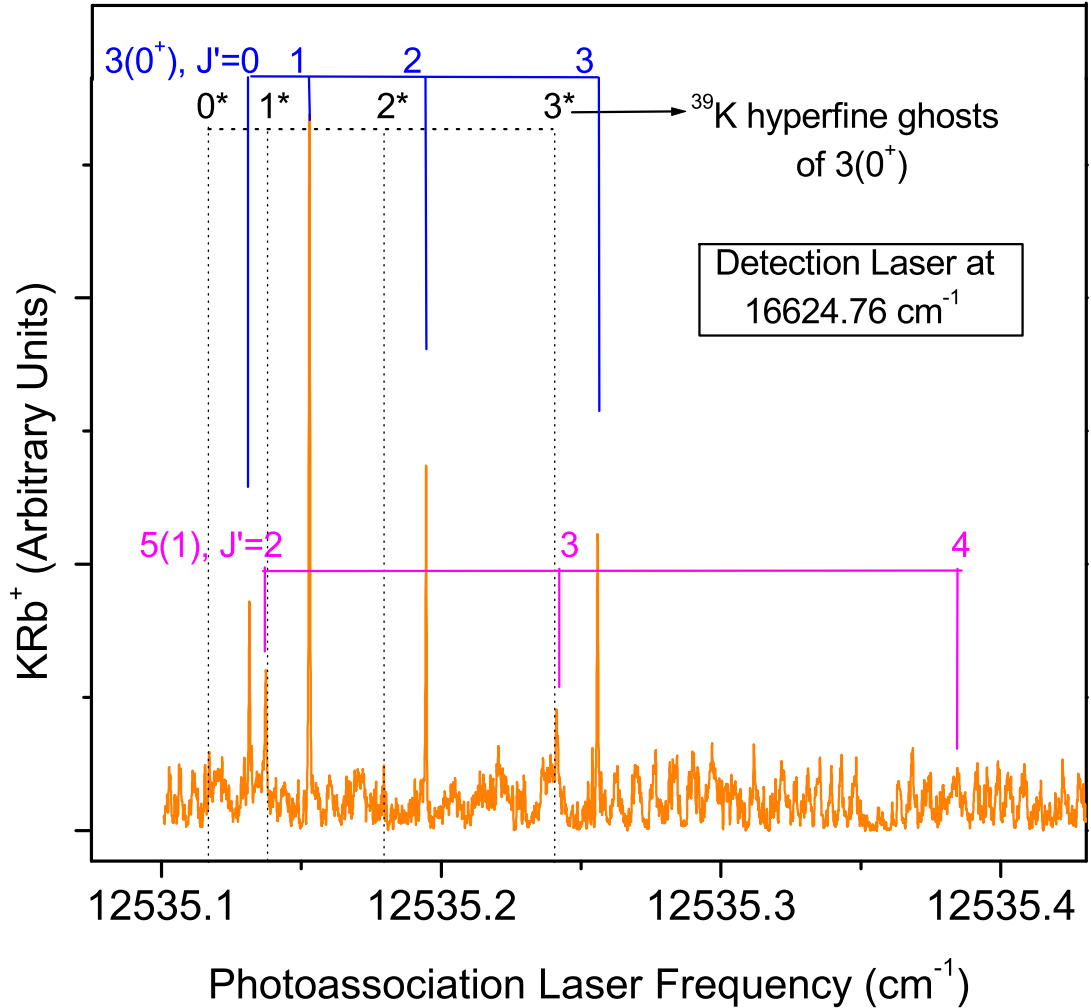}
\caption{\protect\label{Fig11} Redrawing of the PA spectra shown in Figure 5b with additional dotted lines indicating the predicted positions of $^{39}$K hyperfine ghosts of the $3(0^+), v'=118$ rotational band.}   
\end{figure}

To investigate further, we acquired a pair of REMPI spectra over the same wavelength region, with PA first to 5(1), $J'=2$ and then to $3(0^+), J'=1$. Figure 12 shows the resulting spectra. If one of the two lines is the hyperfine ghost of the other, then their REMPI spectra should be almost identical. However, Figure 12 clearly shows that the two REMPI spectra are quite different. 

\begin{figure}[h]
\centering\vskip 0mm
\includegraphics[clip, width=\linewidth]{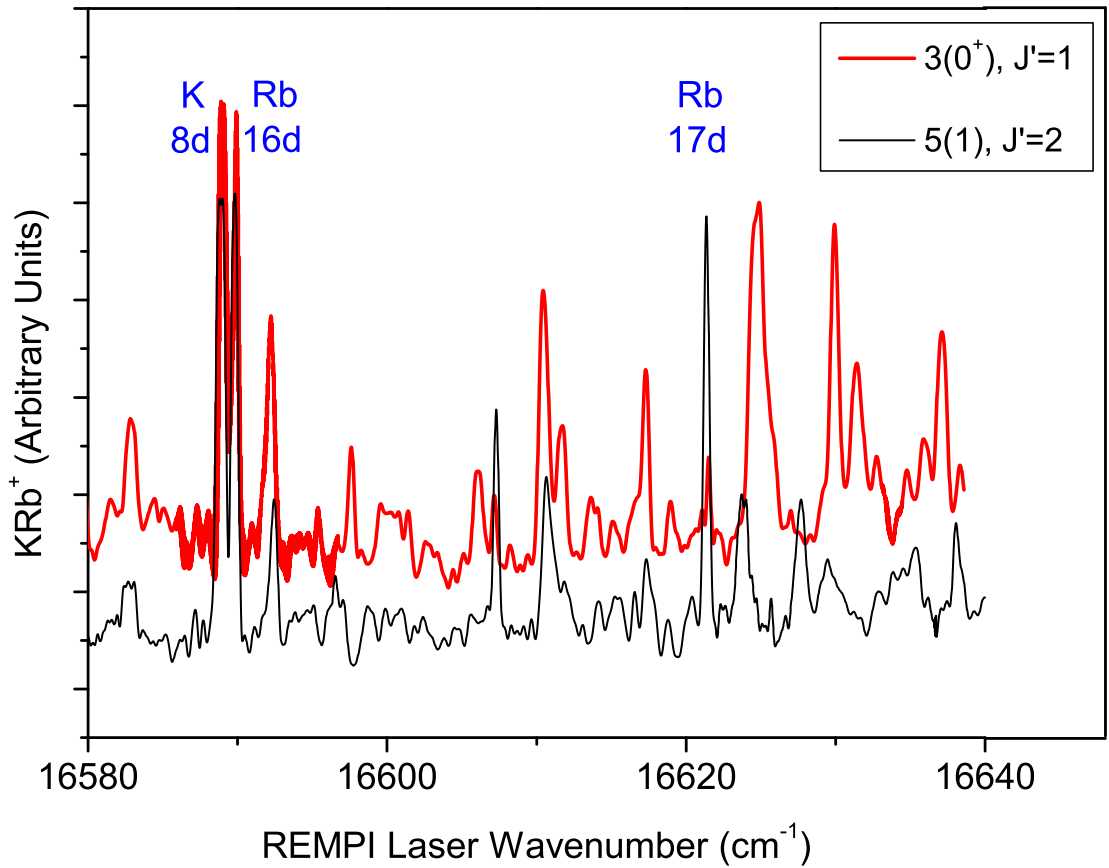}
\caption{\protect\label{Fig12} Comparison of the REMPI spectra with PA fixed at 5(1), $J'=2$ (black, narrower width) and at $3(0^+), J'=1$ (red, thicker width). The $``nd$'' notations indicates the atomic two-photon transitions that leaked into the molecular channel. These lines serve as precise internal references for calibration of each spectrum.}   
\end{figure}

Hence, we conclude that our assignments of the rotational band corresponding to 5(1), $v'=17$ (refer to Figure 5) is correct and they are not hyperfine ghosts of the rotational bands of $3(0^+), v'=118$.
%
%

\begin{thebibliography}{23}%
\makeatletter
\providecommand \@ifxundefined [1]{%
 \@ifx{#1\undefined}
}%
\providecommand \@ifnum [1]{%
 \ifnum #1\expandafter \@firstoftwo
 \else \expandafter \@secondoftwo
 \fi
}%
\providecommand \@ifx [1]{%
 \ifx #1\expandafter \@firstoftwo
 \else \expandafter \@secondoftwo
 \fi
}%
\providecommand \natexlab [1]{#1}%
\providecommand \enquote  [1]{``#1''}%
\providecommand \bibnamefont  [1]{#1}%
\providecommand \bibfnamefont [1]{#1}%
\providecommand \citenamefont [1]{#1}%
\providecommand \href@noop [0]{\@secondoftwo}%
\providecommand \href [0]{\begingroup \@sanitize@url \@href}%
\providecommand \@href[1]{\@@startlink{#1}\@@href}%
\providecommand \@@href[1]{\endgroup#1\@@endlink}%
\providecommand \@sanitize@url [0]{\catcode `\\12\catcode `\$12\catcode
  `\&12\catcode `\#12\catcode `\^12\catcode `\_12\catcode `\%12\relax}%
\providecommand \@@startlink[1]{}%
\providecommand \@@endlink[0]{}%
\providecommand \url  [0]{\begingroup\@sanitize@url \@url }%
\providecommand \@url [1]{\endgroup\@href {#1}{\urlprefix }}%
\providecommand \urlprefix  [0]{URL }%
\providecommand \Eprint [0]{\href }%
\providecommand \doibase [0]{http://dx.doi.org/}%
\providecommand \selectlanguage [0]{\@gobble}%
\providecommand \bibinfo  [0]{\@secondoftwo}%
\providecommand \bibfield  [0]{\@secondoftwo}%
\providecommand \translation [1]{[#1]}%
\providecommand \BibitemOpen [0]{}%
\providecommand \bibitemStop [0]{}%
\providecommand \bibitemNoStop [0]{.\EOS\space}%
\providecommand \EOS [0]{\spacefactor3000\relax}%
\providecommand \BibitemShut  [1]{\csname bibitem#1\endcsname}%
\let\auto@bib@innerbib\@empty
\bibitem [{\citenamefont {{W.C. Stwalley}}\ \emph {et~al.}()\citenamefont
  {{W.C. Stwalley}}, \citenamefont {{P.L. Gould}}, \citenamefont {{and E.E.
  Eyler}}, \citenamefont {{in R.V. Krems}}, \citenamefont {{W.C. Stwalley}},\
  and\ \citenamefont {{B. Friedrich, Editors}}}]{Book09}%
  \BibitemOpen
  \bibfield  {author} {\bibinfo {author} {\bibnamefont {{W.C. Stwalley}}},
  \bibinfo {author} {\bibnamefont {{P.L. Gould}}}, \bibinfo {author}
  {\bibnamefont {{and E.E. Eyler}}}, \bibinfo {author} {\bibnamefont {{in R.V.
  Krems}}}, \bibinfo {author} {\bibnamefont {{W.C. Stwalley}}}, \ and\ \bibinfo
  {author} {\bibnamefont {{B. Friedrich, Editors}}},\ }\href@noop {} {\enquote
  {\bibinfo {title} {Cold molecules: Theory, experiment, applications},}\
  }\bibinfo {howpublished} {(Taylor and Francis, NY, 2009)},\ \bibinfo {note}
  {page 169}\BibitemShut {NoStop}%
\bibitem [{\citenamefont {Wang}\ \emph
  {et~al.}(2004{\natexlab{a}})\citenamefont {Wang}, \citenamefont {Qi},
  \citenamefont {Stone}, \citenamefont {Nikolayeva}, \citenamefont {Wang},
  \citenamefont {Hattaway}, \citenamefont {Gensemer}, \citenamefont {Gould},
  \citenamefont {Eyler},\ and\ \citenamefont {Stwalley}}]{Wang04a}%
  \BibitemOpen
  \bibfield  {author} {\bibinfo {author} {\bibfnamefont {D.}~\bibnamefont
  {Wang}}, \bibinfo {author} {\bibfnamefont {J.}~\bibnamefont {Qi}}, \bibinfo
  {author} {\bibfnamefont {M.~F.}\ \bibnamefont {Stone}}, \bibinfo {author}
  {\bibfnamefont {O.}~\bibnamefont {Nikolayeva}}, \bibinfo {author}
  {\bibfnamefont {H.}~\bibnamefont {Wang}}, \bibinfo {author} {\bibfnamefont
  {B.}~\bibnamefont {Hattaway}}, \bibinfo {author} {\bibfnamefont {S.~D.}\
  \bibnamefont {Gensemer}}, \bibinfo {author} {\bibfnamefont {P.~L.}\
  \bibnamefont {Gould}}, \bibinfo {author} {\bibfnamefont {E.~E.}\ \bibnamefont
  {Eyler}}, \ and\ \bibinfo {author} {\bibfnamefont {W.~C.}\ \bibnamefont
  {Stwalley}},\ }\href@noop {} {\bibfield  {journal} {\bibinfo  {journal}
  {Phys. Rev. Lett.}\ }\textbf {\bibinfo {volume} {93}},\ \bibinfo {pages}
  {243005} (\bibinfo {year} {2004}{\natexlab{a}})}\BibitemShut {NoStop}%
\bibitem [{\citenamefont {Deiglmayr}\ \emph {et~al.}(2008)\citenamefont
  {Deiglmayr}, \citenamefont {Grochola}, \citenamefont {Repp}, \citenamefont
  {M\"ortlbauer}, \citenamefont {Gl\"uck}, \citenamefont {Lange}, \citenamefont
  {Dulieu}, \citenamefont {Wester},\ and\ \citenamefont
  {Weidem\"uller}}]{Deiglmayr08}%
  \BibitemOpen
  \bibfield  {author} {\bibinfo {author} {\bibfnamefont {J.}~\bibnamefont
  {Deiglmayr}}, \bibinfo {author} {\bibfnamefont {A.}~\bibnamefont {Grochola}},
  \bibinfo {author} {\bibfnamefont {M.}~\bibnamefont {Repp}}, \bibinfo {author}
  {\bibfnamefont {K.}~\bibnamefont {M\"ortlbauer}}, \bibinfo {author}
  {\bibfnamefont {C.}~\bibnamefont {Gl\"uck}}, \bibinfo {author} {\bibfnamefont
  {J.}~\bibnamefont {Lange}}, \bibinfo {author} {\bibfnamefont
  {O.}~\bibnamefont {Dulieu}}, \bibinfo {author} {\bibfnamefont
  {R.}~\bibnamefont {Wester}}, \ and\ \bibinfo {author} {\bibfnamefont
  {M.}~\bibnamefont {Weidem\"uller}},\ }\href {\doibase
  10.1103/PhysRevLett.101.133004} {\bibfield  {journal} {\bibinfo  {journal}
  {Phys. Rev. Lett.}\ }\textbf {\bibinfo {volume} {101}},\ \bibinfo {pages}
  {133004} (\bibinfo {year} {2008})}\BibitemShut {NoStop}%
\bibitem [{\citenamefont {Zabawa}\ \emph {et~al.}(2011)\citenamefont {Zabawa},
  \citenamefont {Wakim}, \citenamefont {Haruza},\ and\ \citenamefont
  {Bigelow}}]{Zabawa11}%
  \BibitemOpen
  \bibfield  {author} {\bibinfo {author} {\bibfnamefont {P.}~\bibnamefont
  {Zabawa}}, \bibinfo {author} {\bibfnamefont {A.}~\bibnamefont {Wakim}},
  \bibinfo {author} {\bibfnamefont {M.}~\bibnamefont {Haruza}}, \ and\ \bibinfo
  {author} {\bibfnamefont {N.~P.}\ \bibnamefont {Bigelow}},\ }\href {\doibase
  10.1103/PhysRevA.84.061401} {\bibfield  {journal} {\bibinfo  {journal} {Phys.
  Rev. A}\ }\textbf {\bibinfo {volume} {84}},\ \bibinfo {pages} {061401}
  (\bibinfo {year} {2011})}\BibitemShut {NoStop}%
\bibitem [{\citenamefont {Aikawa}\ \emph {et~al.}(2010)\citenamefont {Aikawa},
  \citenamefont {Akamatsu}, \citenamefont {Hayashi}, \citenamefont {Oasa},
  \citenamefont {Kobayashi}, \citenamefont {Naidon}, \citenamefont {Kishimoto},
  \citenamefont {Ueda},\ and\ \citenamefont {Inouye}}]{Aikawa10}%
  \BibitemOpen
  \bibfield  {author} {\bibinfo {author} {\bibfnamefont {K.}~\bibnamefont
  {Aikawa}}, \bibinfo {author} {\bibfnamefont {D.}~\bibnamefont {Akamatsu}},
  \bibinfo {author} {\bibfnamefont {M.}~\bibnamefont {Hayashi}}, \bibinfo
  {author} {\bibfnamefont {K.}~\bibnamefont {Oasa}}, \bibinfo {author}
  {\bibfnamefont {J.}~\bibnamefont {Kobayashi}}, \bibinfo {author}
  {\bibfnamefont {P.}~\bibnamefont {Naidon}}, \bibinfo {author} {\bibfnamefont
  {T.}~\bibnamefont {Kishimoto}}, \bibinfo {author} {\bibfnamefont
  {M.}~\bibnamefont {Ueda}}, \ and\ \bibinfo {author} {\bibfnamefont
  {S.}~\bibnamefont {Inouye}},\ }\href {\doibase
  10.1103/PhysRevLett.105.203001} {\bibfield  {journal} {\bibinfo  {journal}
  {Phys. Rev. Lett.}\ }\textbf {\bibinfo {volume} {105}},\ \bibinfo {pages}
  {203001} (\bibinfo {year} {2010})}\BibitemShut {NoStop}%
\bibitem [{\citenamefont {Ospelkaus}\ \emph {et~al.}(2010)\citenamefont
  {Ospelkaus}, \citenamefont {Ni}, \citenamefont {Qu\'em\'ener}, \citenamefont
  {Neyenhuis}, \citenamefont {Wang}, \citenamefont {de~Miranda}, \citenamefont
  {Bohn}, \citenamefont {Ye},\ and\ \citenamefont {Jin}}]{Ospelkaus10}%
  \BibitemOpen
  \bibfield  {author} {\bibinfo {author} {\bibfnamefont {S.}~\bibnamefont
  {Ospelkaus}}, \bibinfo {author} {\bibfnamefont {K.-K.}\ \bibnamefont {Ni}},
  \bibinfo {author} {\bibfnamefont {G.}~\bibnamefont {Qu\'em\'ener}}, \bibinfo
  {author} {\bibfnamefont {B.}~\bibnamefont {Neyenhuis}}, \bibinfo {author}
  {\bibfnamefont {D.}~\bibnamefont {Wang}}, \bibinfo {author} {\bibfnamefont
  {M.~H.~G.}\ \bibnamefont {de~Miranda}}, \bibinfo {author} {\bibfnamefont
  {J.~L.}\ \bibnamefont {Bohn}}, \bibinfo {author} {\bibfnamefont
  {J.}~\bibnamefont {Ye}}, \ and\ \bibinfo {author} {\bibfnamefont {D.~S.}\
  \bibnamefont {Jin}},\ }\href {\doibase 10.1103/PhysRevLett.104.030402}
  {\bibfield  {journal} {\bibinfo  {journal} {Phys. Rev. Lett.}\ }\textbf
  {\bibinfo {volume} {104}},\ \bibinfo {pages} {030402} (\bibinfo {year}
  {2010})}\BibitemShut {NoStop}%
\bibitem [{\citenamefont {Stwalley}\ \emph {et~al.}(2010)\citenamefont
  {Stwalley}, \citenamefont {Banerjee}, \citenamefont {Bellos}, \citenamefont
  {Carollo}, \citenamefont {Recore},\ and\ \citenamefont
  {Mastroianni}}]{Banerjee10}%
  \BibitemOpen
  \bibfield  {author} {\bibinfo {author} {\bibfnamefont {W.~C.}\ \bibnamefont
  {Stwalley}}, \bibinfo {author} {\bibfnamefont {J.}~\bibnamefont {Banerjee}},
  \bibinfo {author} {\bibfnamefont {M.}~\bibnamefont {Bellos}}, \bibinfo
  {author} {\bibfnamefont {R.}~\bibnamefont {Carollo}}, \bibinfo {author}
  {\bibfnamefont {M.}~\bibnamefont {Recore}}, \ and\ \bibinfo {author}
  {\bibfnamefont {M.}~\bibnamefont {Mastroianni}},\ }\href {\doibase
  10.1021/jp901803f} {\bibfield  {journal} {\bibinfo  {journal} {J. Phys. Chem
  A}\ }\textbf {\bibinfo {volume} {114}},\ \bibinfo {pages} {81} (\bibinfo
  {year} {2010})}\BibitemShut {NoStop}%
\bibitem [{\citenamefont {Wang}\ \emph
  {et~al.}(2004{\natexlab{b}})\citenamefont {Wang}, \citenamefont {Qi},
  \citenamefont {Stone}, \citenamefont {Nikolayeva}, \citenamefont {Hattaway},
  \citenamefont {Gensemer}, \citenamefont {Wang}, \citenamefont {Zemke},
  \citenamefont {Gould}, \citenamefont {Eyler},\ and\ \citenamefont
  {Stwalley}}]{Wang04b}%
  \BibitemOpen
  \bibfield  {author} {\bibinfo {author} {\bibfnamefont {D.}~\bibnamefont
  {Wang}}, \bibinfo {author} {\bibfnamefont {J.}~\bibnamefont {Qi}}, \bibinfo
  {author} {\bibfnamefont {M.~F.}\ \bibnamefont {Stone}}, \bibinfo {author}
  {\bibfnamefont {O.}~\bibnamefont {Nikolayeva}}, \bibinfo {author}
  {\bibfnamefont {B.}~\bibnamefont {Hattaway}}, \bibinfo {author}
  {\bibfnamefont {S.~D.}\ \bibnamefont {Gensemer}}, \bibinfo {author}
  {\bibfnamefont {H.}~\bibnamefont {Wang}}, \bibinfo {author} {\bibfnamefont
  {W.}~\bibnamefont {Zemke}}, \bibinfo {author} {\bibfnamefont {P.~L.}\
  \bibnamefont {Gould}}, \bibinfo {author} {\bibfnamefont {E.~E.}\ \bibnamefont
  {Eyler}}, \ and\ \bibinfo {author} {\bibfnamefont {W.~C.}\ \bibnamefont
  {Stwalley}},\ }\href@noop {} {\bibfield  {journal} {\bibinfo  {journal} {Eur.
  Phys. J. D}\ }\textbf {\bibinfo {volume} {31}},\ \bibinfo {pages} {165}
  (\bibinfo {year} {2004}{\natexlab{b}})}\BibitemShut {NoStop}%
\bibitem [{\citenamefont {Wang}\ \emph {et~al.}(2005)\citenamefont {Wang},
  \citenamefont {Eyler}, \citenamefont {Gould},\ and\ \citenamefont
  {Stwalley}}]{Wang05}%
  \BibitemOpen
  \bibfield  {author} {\bibinfo {author} {\bibfnamefont {D.}~\bibnamefont
  {Wang}}, \bibinfo {author} {\bibfnamefont {E.~E.}\ \bibnamefont {Eyler}},
  \bibinfo {author} {\bibfnamefont {P.}~\bibnamefont {Gould}}, \ and\ \bibinfo
  {author} {\bibfnamefont {W.~C.}\ \bibnamefont {Stwalley}},\ }\href@noop {}
  {\bibfield  {journal} {\bibinfo  {journal} {Phys. Rev. A}\ }\textbf {\bibinfo
  {volume} {72}},\ \bibinfo {pages} {032052} (\bibinfo {year}
  {2005})}\BibitemShut {NoStop}%
\bibitem [{\citenamefont {Wang}\ \emph {et~al.}(2006)\citenamefont {Wang},
  \citenamefont {Eyler}, \citenamefont {Gould},\ and\ \citenamefont
  {Stwalley}}]{Wang06}%
  \BibitemOpen
  \bibfield  {author} {\bibinfo {author} {\bibfnamefont {D.}~\bibnamefont
  {Wang}}, \bibinfo {author} {\bibfnamefont {E.~E.}\ \bibnamefont {Eyler}},
  \bibinfo {author} {\bibfnamefont {P.~L.}\ \bibnamefont {Gould}}, \ and\
  \bibinfo {author} {\bibfnamefont {W.~C.}\ \bibnamefont {Stwalley}},\ }\href
  {http://stacks.iop.org/0953-4075/39/i=19/a=S03} {\bibfield  {journal}
  {\bibinfo  {journal} {Journal of Physics B: Atomic, Molecular and Optical
  Physics}\ }\textbf {\bibinfo {volume} {39}},\ \bibinfo {pages} {S849}
  (\bibinfo {year} {2006})}\BibitemShut {NoStop}%
\bibitem [{\citenamefont {Wang}\ \emph {et~al.}(2007)\citenamefont {Wang},
  \citenamefont {Kim}, \citenamefont {Ashbaugh}, \citenamefont {Eyler},
  \citenamefont {Gould},\ and\ \citenamefont {Stwalley}}]{Wang07}%
  \BibitemOpen
  \bibfield  {author} {\bibinfo {author} {\bibfnamefont {D.}~\bibnamefont
  {Wang}}, \bibinfo {author} {\bibfnamefont {J.}~\bibnamefont {Kim}}, \bibinfo
  {author} {\bibfnamefont {C.}~\bibnamefont {Ashbaugh}}, \bibinfo {author}
  {\bibfnamefont {E.~E.}\ \bibnamefont {Eyler}}, \bibinfo {author}
  {\bibfnamefont {P.}~\bibnamefont {Gould}}, \ and\ \bibinfo {author}
  {\bibfnamefont {W.~C.}\ \bibnamefont {Stwalley}},\ }\href@noop {} {\bibfield
  {journal} {\bibinfo  {journal} {Phys. Rev. A}\ }\textbf {\bibinfo {volume}
  {75}},\ \bibinfo {pages} {032511} (\bibinfo {year} {2007})}\BibitemShut
  {NoStop}%
\bibitem [{\citenamefont {Banerjee}\ \emph {et~al.}(2012)\citenamefont
  {Banerjee}, \citenamefont {Rahmlow}, \citenamefont {Carollo}, \citenamefont
  {Bellos}, \citenamefont {Eyler}, \citenamefont {Gould},\ and\ \citenamefont
  {Stwalley}}]{Banerjee12}%
  \BibitemOpen
  \bibfield  {author} {\bibinfo {author} {\bibfnamefont {J.}~\bibnamefont
  {Banerjee}}, \bibinfo {author} {\bibfnamefont {D.}~\bibnamefont {Rahmlow}},
  \bibinfo {author} {\bibfnamefont {R.}~\bibnamefont {Carollo}}, \bibinfo
  {author} {\bibfnamefont {M.}~\bibnamefont {Bellos}}, \bibinfo {author}
  {\bibfnamefont {E.~E.}\ \bibnamefont {Eyler}}, \bibinfo {author}
  {\bibfnamefont {P.~L.}\ \bibnamefont {Gould}}, \ and\ \bibinfo {author}
  {\bibfnamefont {W.~C.}\ \bibnamefont {Stwalley}},\ }\href {\doibase
  10.1103/PhysRevA.86.053428} {\bibfield  {journal} {\bibinfo  {journal} {Phys.
  Rev. A}\ }\textbf {\bibinfo {volume} {86}},\ \bibinfo {pages} {053428}
  (\bibinfo {year} {2012})}\BibitemShut {NoStop}%
\bibitem [{\citenamefont {Kim}\ \emph {et~al.}(2009)\citenamefont {Kim},
  \citenamefont {Wang}, \citenamefont {Eyler}, \citenamefont {Gould},\ and\
  \citenamefont {Stwalley}}]{JTKim09}%
  \BibitemOpen
  \bibfield  {author} {\bibinfo {author} {\bibfnamefont {J.-T.}\ \bibnamefont
  {Kim}}, \bibinfo {author} {\bibfnamefont {D.}~\bibnamefont {Wang}}, \bibinfo
  {author} {\bibfnamefont {E.~E.}\ \bibnamefont {Eyler}}, \bibinfo {author}
  {\bibfnamefont {P.~L.}\ \bibnamefont {Gould}}, \ and\ \bibinfo {author}
  {\bibfnamefont {W.~C.}\ \bibnamefont {Stwalley}},\ }\href@noop {} {\bibfield
  {journal} {\bibinfo  {journal} {New J. Phys.}\ }\textbf {\bibinfo {volume}
  {11}},\ \bibinfo {pages} {055020} (\bibinfo {year} {2009})}\BibitemShut
  {NoStop}%
\bibitem [{\citenamefont {Kim}\ \emph {et~al.}(2011{\natexlab{a}})\citenamefont
  {Kim}, \citenamefont {Lee}, \citenamefont {Kim}, \citenamefont {Wang},
  \citenamefont {Stwalley}, \citenamefont {Gould},\ and\ \citenamefont
  {Eyler}}]{JTKim11}%
  \BibitemOpen
  \bibfield  {author} {\bibinfo {author} {\bibfnamefont {J.-T.}\ \bibnamefont
  {Kim}}, \bibinfo {author} {\bibfnamefont {Y.}~\bibnamefont {Lee}}, \bibinfo
  {author} {\bibfnamefont {B.}~\bibnamefont {Kim}}, \bibinfo {author}
  {\bibfnamefont {D.}~\bibnamefont {Wang}}, \bibinfo {author} {\bibfnamefont
  {W.~C.}\ \bibnamefont {Stwalley}}, \bibinfo {author} {\bibfnamefont {P.~L.}\
  \bibnamefont {Gould}}, \ and\ \bibinfo {author} {\bibfnamefont {E.~E.}\
  \bibnamefont {Eyler}},\ }\href {\doibase 10.1039/C1CP21207A} {\bibfield
  {journal} {\bibinfo  {journal} {Phys. Chem. Chem. Phys.}\ }\textbf {\bibinfo
  {volume} {13}},\ \bibinfo {pages} {18755} (\bibinfo {year}
  {2011}{\natexlab{a}})}\BibitemShut {NoStop}%
\bibitem [{\citenamefont {Kim}\ \emph {et~al.}(2011{\natexlab{b}})\citenamefont
  {Kim}, \citenamefont {Lee}, \citenamefont {Kim}, \citenamefont {Wang},
  \citenamefont {Stwalley}, \citenamefont {Gould},\ and\ \citenamefont
  {Eyler}}]{JTKimPRA11}%
  \BibitemOpen
  \bibfield  {author} {\bibinfo {author} {\bibfnamefont {J.-T.}\ \bibnamefont
  {Kim}}, \bibinfo {author} {\bibfnamefont {Y.}~\bibnamefont {Lee}}, \bibinfo
  {author} {\bibfnamefont {B.}~\bibnamefont {Kim}}, \bibinfo {author}
  {\bibfnamefont {D.}~\bibnamefont {Wang}}, \bibinfo {author} {\bibfnamefont
  {W.~C.}\ \bibnamefont {Stwalley}}, \bibinfo {author} {\bibfnamefont {P.~L.}\
  \bibnamefont {Gould}}, \ and\ \bibinfo {author} {\bibfnamefont {E.~E.}\
  \bibnamefont {Eyler}},\ }\href {\doibase 10.1103/PhysRevA.84.062511}
  {\bibfield  {journal} {\bibinfo  {journal} {Phys. Rev. A}\ }\textbf {\bibinfo
  {volume} {84}},\ \bibinfo {pages} {062511} (\bibinfo {year}
  {2011}{\natexlab{b}})}\BibitemShut {NoStop}%
\bibitem [{\citenamefont {Kim}\ \emph {et~al.}()\citenamefont {Kim},
  \citenamefont {Lee}, \citenamefont {Kim}, \citenamefont {Wang}, \citenamefont
  {Stwalley}, \citenamefont {Gould},\ and\ \citenamefont {Eyler}}]{JTKim12}%
  \BibitemOpen
  \bibfield  {author} {\bibinfo {author} {\bibfnamefont {J.-T.}\ \bibnamefont
  {Kim}}, \bibinfo {author} {\bibfnamefont {Y.}~\bibnamefont {Lee}}, \bibinfo
  {author} {\bibfnamefont {B.}~\bibnamefont {Kim}}, \bibinfo {author}
  {\bibfnamefont {D.}~\bibnamefont {Wang}}, \bibinfo {author} {\bibfnamefont
  {W.~C.}\ \bibnamefont {Stwalley}}, \bibinfo {author} {\bibfnamefont {P.~L.}\
  \bibnamefont {Gould}}, \ and\ \bibinfo {author} {\bibfnamefont {E.~E.}\
  \bibnamefont {Eyler}},\ }\href@noop {} {\enquote {\bibinfo {title}
  {Spectroscopic investigation of the $\mathrm{A}$ and $3\,^1\mathrm{\Sigma}^+$
  states of $^{39}\mathrm{K}^{85}\mathrm{Rb}$},}\ }\bibinfo {note} {J. Chem.
  Phys., in press}\BibitemShut {NoStop}%
\bibitem [{\citenamefont {Ketterle}\ \emph {et~al.}(1993)\citenamefont
  {Ketterle}, \citenamefont {Davis}, \citenamefont {Joffe}, \citenamefont
  {Martin},\ and\ \citenamefont {Pritchard}}]{Ketterle93}%
  \BibitemOpen
  \bibfield  {author} {\bibinfo {author} {\bibfnamefont {W.}~\bibnamefont
  {Ketterle}}, \bibinfo {author} {\bibfnamefont {K.~B.}\ \bibnamefont {Davis}},
  \bibinfo {author} {\bibfnamefont {M.~A.}\ \bibnamefont {Joffe}}, \bibinfo
  {author} {\bibfnamefont {A.}~\bibnamefont {Martin}}, \ and\ \bibinfo {author}
  {\bibfnamefont {D.~E.}\ \bibnamefont {Pritchard}},\ }\href {\doibase
  10.1103/PhysRevLett.70.2253} {\bibfield  {journal} {\bibinfo  {journal}
  {Phys. Rev. Lett.}\ }\textbf {\bibinfo {volume} {70}},\ \bibinfo {pages}
  {2253} (\bibinfo {year} {1993})}\BibitemShut {NoStop}%
\bibitem [{\citenamefont {Rousseau}, \citenamefont {Allouche},\ and\
  \citenamefont {Aubert-Fr\'{e}con}(2000)}]{Rousseau00}%
  \BibitemOpen
  \bibfield  {author} {\bibinfo {author} {\bibfnamefont {S.}~\bibnamefont
  {Rousseau}}, \bibinfo {author} {\bibfnamefont {A.~R.}\ \bibnamefont
  {Allouche}}, \ and\ \bibinfo {author} {\bibfnamefont {M.}~\bibnamefont
  {Aubert-Fr\'{e}con}},\ }\href@noop {} {\bibfield  {journal} {\bibinfo
  {journal} {J. Mol. Spectrosc.}\ }\textbf {\bibinfo {volume} {203}},\ \bibinfo
  {pages} {235} (\bibinfo {year} {2000})}\BibitemShut {NoStop}%
\bibitem [{\citenamefont {Pashov}\ \emph {et~al.}(2007)\citenamefont {Pashov},
  \citenamefont {Docenko}, \citenamefont {Tamanis}, \citenamefont {Ferber},
  \citenamefont {Kn\"ockel},\ and\ \citenamefont {Tiemann}}]{Pashov07}%
  \BibitemOpen
  \bibfield  {author} {\bibinfo {author} {\bibfnamefont {A.}~\bibnamefont
  {Pashov}}, \bibinfo {author} {\bibfnamefont {O.}~\bibnamefont {Docenko}},
  \bibinfo {author} {\bibfnamefont {M.}~\bibnamefont {Tamanis}}, \bibinfo
  {author} {\bibfnamefont {R.}~\bibnamefont {Ferber}}, \bibinfo {author}
  {\bibfnamefont {H.}~\bibnamefont {Kn\"ockel}}, \ and\ \bibinfo {author}
  {\bibfnamefont {E.}~\bibnamefont {Tiemann}},\ }\href {\doibase
  10.1103/PhysRevA.76.022511} {\bibfield  {journal} {\bibinfo  {journal} {Phys.
  Rev. A}\ }\textbf {\bibinfo {volume} {76}},\ \bibinfo {pages} {022511}
  (\bibinfo {year} {2007})}\BibitemShut {NoStop}%
\bibitem [{Note1()}]{Note1}%
  \BibitemOpen
  \bibinfo {note} {See supplementary material at [URL will be inserted by AIP]
  for the listing of the four potential energy curves that are derived from the
  \protect \emph {ab initio} potentials.}\BibitemShut {Stop}%
\bibitem [{\citenamefont {Kasahara}\ \emph {et~al.}(1999)\citenamefont
  {Kasahara}, \citenamefont {Fujiwara}, \citenamefont {Okada}, \citenamefont
  {Kato},\ and\ \citenamefont {Baba}}]{Kasahara99}%
  \BibitemOpen
  \bibfield  {author} {\bibinfo {author} {\bibfnamefont {S.}~\bibnamefont
  {Kasahara}}, \bibinfo {author} {\bibfnamefont {C.}~\bibnamefont {Fujiwara}},
  \bibinfo {author} {\bibfnamefont {N.}~\bibnamefont {Okada}}, \bibinfo
  {author} {\bibfnamefont {H.}~\bibnamefont {Kato}}, \ and\ \bibinfo {author}
  {\bibfnamefont {M.}~\bibnamefont {Baba}},\ }\href {\doibase 10.1063/1.480256}
  {\bibfield  {journal} {\bibinfo  {journal} {J. Chem. Phys}\ }\textbf
  {\bibinfo {volume} {111}},\ \bibinfo {pages} {8857} (\bibinfo {year}
  {1999})}\BibitemShut {NoStop}%
\bibitem [{\citenamefont {LeRoy}()}]{LEV74}%
  \BibitemOpen
  \bibfield  {author} {\bibinfo {author} {\bibfnamefont {R.~J.}\ \bibnamefont
  {LeRoy}},\ }\href@noop {} {\enquote {\bibinfo {title} {Level 8.0},}\
  }\bibinfo {howpublished} {University of Waterloo Chemical Physics Research
  Report CP-642R},\ \bibinfo {note} {web site
  http://leroy.uwaterloo.ca}\BibitemShut {NoStop}%
\bibitem [{\citenamefont {Herzberg}(1989)}]{Herzberg}%
  \BibitemOpen
  \bibfield  {author} {\bibinfo {author} {\bibfnamefont {G.}~\bibnamefont
  {Herzberg}},\ }\href@noop {} {\emph {\bibinfo {title} {Molecular spectra and
  molecular structure I, Spectra of diatomic molecules}}}\ (\bibinfo
  {publisher} {Robert E. Krieger publishing Co.},\ \bibinfo {address} {Malabar,
  Florida},\ \bibinfo {year} {1989})\BibitemShut {NoStop}%
\end{thebibliography}

%

\end{document}